\documentclass[11pt,twocolumn]{article}

\usepackage[top= 2cm,bottom=2cm,left=2.00cm,right=2.00cm]{geometry}

\usepackage{graphicx}
\usepackage{amsmath,amssymb,amsfonts}
\usepackage{amsthm}
\usepackage{mathrsfs}
\usepackage{xcolor}
\usepackage{textcomp}
\usepackage{booktabs}
\usepackage{multirow}
\usepackage{algorithm}
\usepackage{algorithmicx}
\usepackage{algpseudocode}
\usepackage{listings}
\usepackage[numbers,sort&compress]{natbib}
\usepackage[colorlinks=true,linkcolor=blue,citecolor=blue,urlcolor=blue]{hyperref}

\renewcommand{\d}{\mathrm{d}}
\newcommand{\e}{\mathrm{e}}
\renewcommand{\i}{\mathrm{i}}
\newcommand{\D}{\mathrm{D}}

\newcommand{\PD}[2]{\frac{\partial #1}{\partial #2}}
\newcommand{\FD}[2]{\frac{\d #1}{\d #2}}

\newcommand{\ZSet}{\mathbb{Z}}
\newcommand{\RSet}{\mathbb{R}}
\newcommand{\CSet}{\mathbb{C}}
\newcommand{\SSet}{\mathbb{S}}

\begin{document}

\title{\vspace*{-1.25cm}Adaptive conduction delays and phase locking in spiking Haken Lighthouse networks}

\author{Stephen Coombes\thanks{Corresponding author:\\ \texttt{stephen.coombes@nottingham.ac.uk}}, R\"udiger Thul, Stefan Ruschel, and Rachel Nicks\\[0.5em]
School of Mathematical Sciences, University of Nottingham, Nottingham, NG7 2RD, UK}

\date{}

\maketitle

\abstract{
We develop a theory of phase-locked activity in delayed spiking networks using the Haken Lighthouse model as an analytically tractable event-based description of neural dynamics. For networks with fixed delays, we derive self-consistency conditions for phase-locked states and an associated linear stability theory formulated directly in terms of spike-time perturbations. The framework is illustrated for a delayed autapse, a reciprocally coupled two-cell network, and spatially structured rings with distance-dependent coupling and conduction delays, where circulant symmetry allows stability to be decomposed into Fourier modes. We then introduce an activity-dependent white matter plasticity rule in which myelination modulates axonal conduction speed and hence communication delay. This leads naturally to a slow--fast system with state-dependent delays, in which frozen phase-locked branches organise the adaptive dynamics. The plasticity rule selects commensurate delay--period relationships, providing a mechanism for the emergence of synchrony, other frequency-locked states, slow switching between competing phase-locked patterns, and the organisation of heterogeneous delays into discrete delay--period classes. Direct simulations of the event-driven network support the analytical predictions and illustrate how adaptive conduction can reshape the attractor structure of a delayed spiking network and generate long-timescale transitions. These results provide a tractable mathematical framework for studying how activity-dependent myelination may regulate temporal coordination, synchrony, and communication through coherence in spiking neural systems.
}

\bigskip


\section{Introduction}

A central goal of theoretical neuroscience is to construct tractable models of spiking neuronal networks. Achieving this requires a single-node model capable of generating discrete events---action potentials---that, when embedded in a synaptic network, can give rise to the collective behaviours observed in biological neural systems. 
Simple spiking neural network models, such as those built from interacting integrate-and-fire units, exhibit rich emergent behaviours but remain notoriously difficult to analyse, especially when one moves beyond regular firing patterns or highly symmetric forms of coupling \cite{Coombes2012}. In contrast, firing-rate descriptions and coupled phase-oscillator models offer greater mathematical tractability, but necessarily average over, or otherwise suppress, some of the event-based structure of neural communication. The Haken Lighthouse model offers a promising bridge between these modelling paradigms. Introduced by Hermann Haken within the wider programme of Synergetics, it describes a network of pulse-coupled units whose firing is determined by the rotation of a phase-like variable driven by synaptic input \cite{Haken2002}. It is therefore spiking, but in a form that preserves enough structure to make the analysis of collective states feasible.

A recent revisiting of the Lighthouse model developed new mathematical tools for analysing self-organisation in spiking networks, including synchrony, waves, bumps, and pattern-forming instabilities \cite{Coombes2025}. That work also reaffirmed the value of the Lighthouse model as a link between spiking descriptions, phase-oscillator ideas, and rate-based neural fields. However, the treatment of delays in that study was limited. This is a significant restriction. Delays are not small technical corrections to neural dynamics: they are fundamental components of communication in nervous systems \cite{Campbell2007}. Axonal propagation, synaptic processing, dendritic filtering, and long-range anatomical embedding all introduce timing constraints, and these timing constraints can shape synchrony, phase locking, travelling activity, multistability, and the response of a network to perturbations.
Recent work on large-scale rate-based brain models has similarly shown that fixed white matter delays can strongly influence synchrony and collective dynamics across spatially distributed neural populations \cite{Sayli2024}.

Delays are also increasingly relevant for NeuroAI and neuromorphic computation. Spiking neural networks are attractive partly because their event-based nature suggests a route toward energy-efficient computation \cite{Ornes2025,DiCaterina2024}. Yet event-based computation is intrinsically temporal. The timing of a spike, and not merely its occurrence, can carry information. In this setting, delays are not just imperfections to be engineered away; they are additional degrees of freedom that can enhance memory, temporal processing, and reservoir-like computation \cite{Tavakoli2024}. A mathematical theory of spiking networks that treats delays as primary dynamical ingredients is therefore needed both for neuroscience and for next-generation brain-inspired computing.

The biological motivation for treating delays seriously is strengthened by the growing evidence that white matter is plastic. Myelination controls axonal conduction speed and hence the delay with which signals arrive at their targets. It is now clear that myelination is activity dependent, with neural activity influencing oligodendrocyte lineage cells and modulating conduction properties through changes in myelin structure on a time-scale that is slow compared to neural dynamics \cite{Gibson2014,Mount2017,Xin2020}. From a dynamical perspective this means that delays should not always be viewed as fixed parameters. Instead, they may themselves evolve slowly in response to the activity that they help organise. Recent modelling work has begun to explore this idea in phase-oscillator and large-scale network settings, showing how adaptive myelination can promote synchronisation, resilience, and slow switching in delayed networks \cite{Pajevic2014,Park2020,Talidou2022,Klinshov2024,Lefebvre2025,Pajevic2023,Ruschel2026,Jolly2026}. The aim of the present paper is to develop an analogous theory directly at the level of a spiking Lighthouse network.

The main contribution of this paper is therefore twofold. First, we develop a framework for the existence and stability of phase-locked states in Haken Lighthouse networks with multiple fixed delays. The theory is formulated in terms of event times and leads naturally to a characteristic equation for perturbations of firing times. Secondly, we use this fixed-delay theory as the backbone for an adaptive model in which conduction speeds, and hence delays, evolve slowly according to a phenomenological rule for white matter plasticity. This produces a state-dependent delay problem in which the fast spiking dynamics and the slow evolution of conduction delays are coupled. The resulting framework gives a tractable way to study how adaptive delays can promote synchrony, organise switching between phase-locked states, and sculpt networks toward commensurate timing classes that preserve phase information along long-range pathways. In this sense, the model connects naturally with the communication-through-coherence viewpoint, in which the efficacy of interaction between neuronal populations depends on the phase at which inputs arrive \cite{Fries2005,Fries2015}.

The structure of the paper is as follows. In Sec.~\ref{Sec:Lighthouse} we introduce the Haken Lighthouse model on a graph, with synaptic interactions represented by causal kernels and communication delays. In Sec.~\ref{Sec:Phaselocked} we derive the equations for phase-locked states with arbitrary network topology and multiple delays, and formulate their linear stability in terms of perturbations of firing times. Section~\ref{Sec:autapse} treats the single-node autapse as a minimal example in which folds of regular-spiking branches and dynamic instabilities of inter-spike intervals can be identified explicitly. In Sec.~\ref{Sec:N=2} we analyse a reciprocal two-node network, including synchronous, anti-synchronous, and asymmetric phase-locked states, and show how the two-delay problem can be reduced to an equivalent one-delay description. Section~\ref{Sec:Ring} treats ring networks with distance-dependent interactions and delays, where circulant structure allows the stability problem for twisted states to be diagonalised. In Sec.~\ref{Sec:WMPlasticity} we introduce a white-matter plasticity rule for the slow evolution of conduction speeds, develop a slow--fast interpretation of the adaptive system, and illustrate how the rule can promote convergence, switching, and commensurate timing. 
Finally, in Sec.~\ref{Sec:Discussion} we summarise the main results and discuss how adaptive conduction delays in the Lighthouse model may provide a useful mathematical framework for studying biological white matter networks, and for understanding how delay-based mechanisms can regulate temporal coordination, synchrony and switching in spiking neural systems.

\section{The Lighthouse model\label{Sec:Lighthouse}}

A recent revisiting of the Lighthouse model showed how its combination of event-driven spiking and synaptic filtering can be exploited to develop a mathematical theory of self-organised activity in spiking networks \cite{Coombes2025}.  In particular, it was shown that the model supports a range of collective states familiar from pattern-formation theory, including synchrony, travelling waves, localised bumps and instabilities of spatially homogeneous activity.  These results position the Lighthouse model as a tractable framework for connecting pulse-coupled spiking dynamics with the geometric and spectral ideas more commonly used in the analysis of rate-based neural networks.  The present paper builds on this viewpoint, but now places conduction delays at the centre of the analysis.

We consider the Haken Lighthouse model on a directed weighted graph with $N$ nodes and connection strengths $w_{ij}\in\RSet$.  The state of node $i$ is described by a phase variable $\theta_i\in\SSet$, driven by a synaptic input $\psi_i\in\RSet$.  The model takes the form
\begin{equation}
\begin{split}
\FD{}{t}\theta_i(t) &= S\left(\psi_i(t)\right),\\
\psi_i(t) &= \sum_{j=1}^N w_{ij}\left(\eta_{ij}\ast s_j\right)(t),\\
s_j(t) &= \sum_{m\in\ZSet}\delta(t-T_j^m),
\end{split}
\label{LHmodel}
\end{equation}
for $i=1,\ldots,N$.  Here, $S$ is a non-negative firing-rate function, $\eta_{ij}$ is a causal synaptic response kernel from node $j$ to node $i$, and $\ast$ denotes temporal convolution.  The spike train $s_j$ is generated by the firing times $T_j^m$, of node $j$ with $m\in\ZSet$.  These firing times are determined by the threshold condition $\theta_j(T_j^m)=2\pi$, with the phase then understood to wrap around the circle.  Thus the Lighthouse model is naturally formulated as an event-driven spiking system, in which the continuous evolution of the phase variables is coupled to the discrete sequence of spike times.

A typical choice for $S$ is the sigmoidal function
\begin{equation}
S(x) = \frac{1}{1+\e^{-\beta(x-h)}},
\label{S}
\end{equation}
where $h\in\RSet$ is a threshold and $\beta>0$ controls the steepness.  In what follows we shall assume more generally that $S$ is positive, monotonically increasing, and at least once differentiable.

Communication delays enter through the synaptic kernels.  In this paper we take $\eta_{ij}(t)=\eta(t-\tau_{ij})$, where $\tau_{ij}\geq 0$ is the propagation delay from node $j$ to node $i$.  For the underlying causal response we use the normalised $\alpha$-function
\begin{equation}
\eta(t)=\alpha^2t\e^{-\alpha t}H(t),
\label{alphafunction}
\end{equation}
where $H$ is the Heaviside function.  The parameter $\alpha$ sets the synaptic time scale, with larger values corresponding to faster synaptic responses.  Since $\eta$ is the Green's function of the linear differential operator
\begin{equation}
Q=\left(1+\frac{1}{\alpha}\frac{\d}{\d t}\right)^2,
\label{Qoperator}
\end{equation}
the synaptic input can equivalently be written in the filtered-spike form
\begin{equation}
Q\psi_i(t)=\sum_{j=1}^N w_{ij}s_j(t-\tau_{ij}).
\label{filteredspikes}
\end{equation}
This representation makes explicit the mechanisms that will be central throughout the paper: spikes are communicated through delayed pathways, and the postsynaptic drive is obtained by filtering these delayed spike trains through a causal synaptic kernel.

\section{Phase-locked states\label{Sec:Phaselocked}}

We now turn to the construction and stability of phase-locked states.  The event-driven nature of the Lighthouse model makes it natural to formulate the problem directly in terms of firing times.  Integrating (\ref{LHmodel}) from one firing event at time $T_i^m$ to the next at time $T_i^{m+1}$ gives
\begin{align}
& \theta_i(T_i^{m+1}) - \theta_i(T_i^{m}) = 2 \pi \nonumber \\
&= 
\int_{0}^{T_i^{m+1}-T_i^m}
\! \! \! \!  \! \! \! \! \! \!  \d t \,
S \left (
\sum_{j=1}^N w_{ij} \! \! \! \! \! \! \! \! \! \!
\sum_{\{ p \mid T_j^p < T_i^{m+1} \}} \! \! \! \! \! \! \! \! \! \! 
\eta(t +T_i^m -T_j^p - \tau_{ij})
\right ) 
\label{LHmap}
\end{align}
This expression is an exact firing-time map: the next spike of node $i$ is determined by the delayed spikes from all presynaptic nodes that arrive before the threshold crossing.

We now define a phase-locked state according to $T_i^m = mT + \phi_i T$ with $\phi_i \in [0,1)$.  In this case (\ref{LHmap}) reduces to the set of equations:
\begin{equation}
2 \pi = \int_{0}^{T} \d t \, S \left ( \sum_{j=1}^N w_{ij} P(t +(\phi_i-\phi_j)T - \tau_{ij}) \right ) , 
\label{Phaselocked}
\end{equation}
for $i=1,\ldots, N$.
Here, the $T$-periodic function $P(t)$ is given by $P(t) = \sum_{m \in \ZSet} \eta(t - m T)$.  
This may be equivalently represented by a Fourier series of the form $P(t) = \sum_n P_n \e^{\i \omega_n t }$ with $\omega_n = 2 \pi n/T$
\begin{equation}
P_n = \frac{1}{T} \widehat{\eta}(\omega_n) , \quad \widehat{\eta}(\omega) = \int_{-\infty}^ \infty \d t \, \eta(t) \e^{-\i \omega t} .
\end{equation}
Here $\widehat{\eta}$ is recognised as the Fourier transform of $\eta$.  
For the $\alpha$-function we have that
\begin{equation}
\widehat{\eta} (\omega) = \frac{1}{(1+\i \omega/\alpha)^2} .
\end{equation}
We note that we may also evaluate $P(t)$ in closed form by recognising it as a geometric progression that may be summed to give
\begin{align}
P(t) &= -\alpha^2 \FD{}{\alpha} \sum_{m \leq 0}  \e^{-\alpha (t-mT)} \nonumber \\
&= \frac{\alpha^2 \e^{-\alpha t}}{1-\e^{-\alpha T}}\left [ t + \frac{T \e^{-\alpha T}}{1-\e^{-\alpha T}} \right ] , \qquad t \in [0, T).
\end{align}

The system of equations (\ref{Phaselocked}) is invariant under the shift $\phi_i \rightarrow \phi_i + a$ for some arbitrary constant $a$.  This highlights a phase-shift symmetry of the system, arising from the fact that the equations defining a phase-locked state only depend upon phase differences.  The same system of equations is also invariant to a constant shift for one (and only one) of the delays, namely $\tau_{ij} \rightarrow \tau_{ij} + a$ for some pair $(i,j)$.  The same is true if a constant is added to \textit{all} delays.
This arises because the integral of a periodic function over its period is invariant with respect to the choice of an origin. A consequence of this is that if there is only one delay in the system then the emergent period is independent of that delay.
We also note that if $S$ is a linear function then (\ref{Phaselocked}) is independent of all delays.
If we fix one phase, say $\phi_1=0$, then (\ref{Phaselocked}) gives a system of $N$ equations in the $N$ unknowns $(T, \phi_2, \ldots, \phi_N)$ (for fixed delays).  Moreover, if it proves convenient we might fix one delay in the system and refer all others to this.  

To determine the linear stability of this phase-locked state we consider a perturbed flow denoted by $\widetilde{\theta}_i(t)  = {\theta}_i(t) + \delta{\theta}_i(t)$ with a corresponding set of perturbed firing times that we denote as $\widetilde{T}_i^m =\widetilde{T}_i^m + \delta \widetilde{T}_i^m$.  This perturbed system can be integrated to yield (\ref{LHmap}) under the replacement $\theta_i \rightarrow \widetilde{\theta}_i$ and $T_i^m \rightarrow \widetilde{T}_i^m$.  A Taylor expansion of this set of equations for small perturbations yields
\begin{align}
0 &\simeq \int_0^T \d t \, S' \left ( \sum_{j=1}^N w_{ij} P(t+ A_{ij}) \right ) \times  \nonumber \\
& \sum_{k=1}^N w_{ik} \sum_{p \in \ZSet} \eta' (t+ pT + A_{ik}) \left ( \delta T_i^m - \delta T_k^{m-p} \right ) \nonumber \\
&+ ( \delta T_i^{m+1} - \delta T_i^{m} ) S \left ( \sum_{j=1}^N w_{ij} P(A_{ij}) \right ) ,
\end{align}
where $A_{ij} = (\phi_i-\phi_j)T -\tau_{ij}$.  This linear difference equation has solutions of the form $\delta T_i^m = \delta T_i \e^{m \lambda}$ with $\lambda \in \CSet$.  These solutions satisfy the system of linear equations
\begin{equation}
\left (\e^{\lambda} - 1 \right ) \dot{\theta}_i \delta T_i + \sum_{j=1}^N \mathcal{L}_{ij} (\lambda) \delta T_j = 0,
\label{eigsystem}
\end{equation}
where $\dot{\theta}_i = S (\sum_j w_{ij} P(A_{ij}))$, $\mathcal{L}_{ij}(\lambda) = -w_{ij}(\lambda) + \delta_{ij} \sum_k w_{ik}(0)$, $w_{ij} (\lambda) = w_{ij} G_{ij} (\lambda)$ and
\begin{align}
G_{ik}(\lambda) &= \sum_{p \in \ZSet} \e^{-p \lambda} \int_0^T \d t \, S' \left ( \sum_{j=1}^N w_{ij} P(t+A_{ij}) \right ) \nonumber \\
& \times \eta'(t+pT + A_{ik}) .
\label{Gik}
\end{align}
Introducing the $N \times N$ diagonal matrix $\dot{\Theta}$ with components $[\dot{\Theta}]_{ij} = \dot{\theta}_i \delta_{ij}$ means that we can write (\ref{eigsystem}) in the more compact form $\mathcal{M}(\lambda) \delta T = 0$, where $\delta T = (\delta T_1, \ldots, \delta T_N)$ and
\begin{equation}
\mathcal{M}(\lambda) = \left (\e^{\lambda} - 1 \right ) \dot{\Theta} + \mathcal{L} (\lambda). 
\label{M}
\end{equation}
For non-trivial solutions we must have that $\mathcal{E}(\lambda) = 0$ where $\mathcal{E}(\lambda) = \det \mathcal{M}(\lambda)$.  A phase-locked state will be stable if $\text{Re} \, \lambda <0$.  We note that $(1,1,\ldots, 1)$ is an eigenvector of $\mathcal{L}(0)$ with eigenvalue zero.  Hence, $\mathcal{E}(0) = 0$.  This is expected since the system has phase shift symmetry.

A computationally useful form for $G_{ij}(\lambda)$ can be obtained by using a Fourier integral representation of $\eta'$ as:
\begin{equation}
\eta'(t) = \frac{1}{2 \pi} \int_{-\infty}^\infty \d k \, (\i k) \widehat{\eta} (k) \e^{\i k t} .
\end{equation}
Using this we find
\begin{equation}
G_{ij}(\lambda) = \frac{1}{T} \sum_{n\in \ZSet} f_{i} (\widetilde{\omega}_n (\lambda)) \,
\widehat{\eta} (\widetilde{\omega}_n (\lambda))   \e^{\i (\widetilde{\omega}_n (\lambda)) A_{ij}}  , \label{Gl}
\end{equation}
where $\widetilde{\omega}_n (\lambda) = \omega_n -\i \lambda/T$ ($\omega_n = 2 \pi n /T$) and 
\begin{equation}
f_i(\omega) = \int_0^T \d t \, S'  ( X_i(t))  (\i \omega)\e^{\i \omega t} .
\label{fi}
\end{equation}
where $X_i(t) = \sum_{j=1}^N w_{ij} P(t+A_{ij})$,
and we have made use of the Dirac comb identity $2 \pi \sum_{n \in \ZSet} \delta(x-2 \pi n) = \sum_{p \in \ZSet} \e^{\i p x}$.  
By writing $Y_i(t)=S'(X_i(t))$ as a Fourier series with Fourier coefficients $Y_{i,n} \in \CSet$ we may evaluate (\ref{fi}) to the computationally useful form
$f_i(\omega_n) = \i \omega_n T \, Y_{i,-n}$.

In practice, the spectral equation $\mathcal{E}(\lambda)=0$ can be evaluated as follows.  For a given phase-locked state, the period $T$ and phases $\phi_i$ are first determined from (\ref{Phaselocked}).  For each trial value of $\lambda$, the quantities $G_{ij}(\lambda)$ are then computed from the Fourier representation (\ref{Gl}), with the infinite sum over $n$ truncated symmetrically to $|n|\leq n_{\max}$.  This truncation is used only to evaluate the entries of the $N\times N$ matrix $\mathcal{M}(\lambda)$; it does not enlarge the dimension of the stability matrix.  The roots of the scalar nonlinear equation $\mathcal{E}(\lambda)=\det\mathcal{M}(\lambda)=0$ are then found directly, or in special cases to determine potential bifurcation points by solving the real and imaginary parts of $\mathcal{E}(\i\omega)=0$.  Thus the Fourier truncation provides a spectral approximation of the delayed synaptic response, whereas the determinant is always taken over the network degrees of freedom.

In the examples below, this general construction will be specialised first to a single neuron with delayed self-feedback, then to a reciprocal two-node network, and finally to circulant ring networks where the spatial symmetry allows further diagonalisation.

\section{A single node with an autapse\label{Sec:autapse}}

We first apply the phase-locking framework to the simplest possible delayed network: a single Haken neuron with recurrent self-feedback. This example is useful for two reasons. First, it provides a minimal setting in which the role of delay can be isolated and understood. Second, it illustrates how the general firing-time stability theory detects both folds of regular-spiking branches and dynamic instabilities associated with modulations of the inter-spike intervals.

We consider a single neuron with two forms of self-coupling: an instantaneous autapse (self synapse) of strength $w_0$ and a delayed autapse of strength $w_\tau$ and delay $\tau$. The instantaneous term breaks the
delay-shift invariance discussed in the previous section, allowing the period of a regular spike train to depend non-trivially on the delay. 
Previous work on (perfect) integrate-and-fire and Hodgkin-Huxley-type neuron models, whose recurrent inputs are delayed versions of their output spike trains, has shown that multi-stability of periodic spike trains is possible \cite{Foss1996}.  
Other work by Klinshov \textit{et al}. has shown that simple phase oscillator and Hodgkin-Huxley models with pulsatile delayed feedback can also exhibit so-called ``jittering", whereby a periodic spike train destabilises to temporal patterns with non-equal interspike intervals \cite{Klinshov2015}.  The Haken autapse displays analogous behaviour,
but here the bifurcation structure can be obtained directly from the event-time formulation.

Following our earlier notation and dropping node labels (as we only consider one Haken neuron here), we write the $m$th inter-spike interval (ISI) as $\Delta^{m} = T^{m+1}-T^m$.  For a periodic spike train we have simply that $\Delta^m = T$ for all $m$.  Denoting the self-delay by $\tau$ and the self-connection strength by $w_\tau$, the period $T = T(\tau)$ is then implicitly defined from equation (\ref{Phaselocked}) as the solution to $\mathcal{H}(T) = 0$, where
\begin{equation}
\mathcal{H}(T) = \int_{0}^{T} \d t \, S \left ( w_0 P(t) + w_\tau P(t - \tau) \right ) - 2 \pi .
\label{Hautapse}
\end{equation}
From the linear stability analysis centred around equation (\ref{M}), the periodic spike-train will be stable if $\text{Re} \, \lambda <0$ where $(\e^{\lambda} - 1)\dot{\theta} - [\mathcal{G}(\lambda) - \mathcal{G}(0)]=0 \equiv \mathcal{E}(\lambda)$, where 
\begin{align}
\mathcal{G}(\lambda) &= \sum_{p \in \ZSet} \e^{-p \lambda} \int_0^T \d t \, S' \left (  w_0 P(t) + w_\tau P(t - \tau) \right ) \nonumber \\
& \times  \left [ w_0 \eta'(t+pT) + w_\tau \eta'(t+pT-\tau) \right ] .
\end{align}
Here, $\dot{\theta} = S(w_0 P(0)+ w_\tau P(-\tau))$.
Expanding the characteristic equation near $\lambda= 0$ shows that a double root occurs when $\mathcal{E}(0)=0=\mathcal{E}'(0)$ giving
$\dot{\theta} - \mathcal{G}'(0)=0$, signalling a bifurcation from one branch of periodic spike-trains to another.  This occurs at a fold in the solution branch for $T=T(\tau)$.  To see this first differentiate (\ref{Hautapse}) with respect to $T$ to find
\begin{equation}
\PD{\mathcal{H}}{T} = \dot{\theta} + \int_0^T \d t \, S' (X) \PD{X}{T} , 
\end{equation}
where $X = w_0 P(t) + w_\tau P(t -\tau)$.
Since
\begin{equation}
\PD{X}{T} = \sum_{p \in \ZSet}  p \left [ w_0 \eta'(t+pT) + w_\tau \eta'(t+pT-\tau) \right ],
\end{equation}
and $\mathcal{G}'(0) = - \int_0^T \d t \, S' (X) \partial X / \partial T$,
it follows that $\partial \mathcal{H} /\partial T  = \dot{\theta} - \mathcal{G}'(0) = \mathcal{E}'(0)$.
Hence, a fold $\partial \mathcal{H} /\partial T = 0$ coincides with a double zero of the characteristic equation.

For $\lambda = \i \omega$ a dynamic instability (signalling the onset of an oscillatory modulation of the ISIs) can occur at some value of $\tau$, and this is defined by the simultaneous solution of
\begin{equation}
\begin{split}
\dot{\theta} \left ( \cos \omega -1 \right ) &= \left (\text{Re} \, \mathcal{G}(\i \omega) - \mathcal{G}(0) \right ) ,\\
\dot{\theta} \sin \omega  &= \text{Im} \, \mathcal{G}(\i \omega) ,
\end{split}
\label{dynamic}
\end{equation}
for the pair $(\tau, \omega)$.
The frequency $\omega$ characterises modulation in spike index rather than physical time. If the base spike train has period $T$, then the modulation repeats after 
$q$ spikes if
\begin{equation}
\frac{\omega}{2 \pi} = \frac{p}{q}, \qquad p,q \in \mathbb{Z}.
\end{equation}
In this case the destabilised solution exhibits a $q$-cycle in the ISIs. In particular, $\omega=\pi$ corresponds to period-doubling with alternating long--short intervals, while $\omega=2\pi/3$ gives a three-ISI repeating pattern. If $\omega/2\pi$ is irrational, the modulation is quasi-periodic.
Thus the onset of instability is naturally associated with ``jittering'' of the spike train, with ISIs of the form $\Delta^m = T + \varepsilon \cos(m\omega + \phi (\omega)) + \cdots$.

\begin{figure}[htbp]
\centering
\includegraphics[width=\columnwidth]{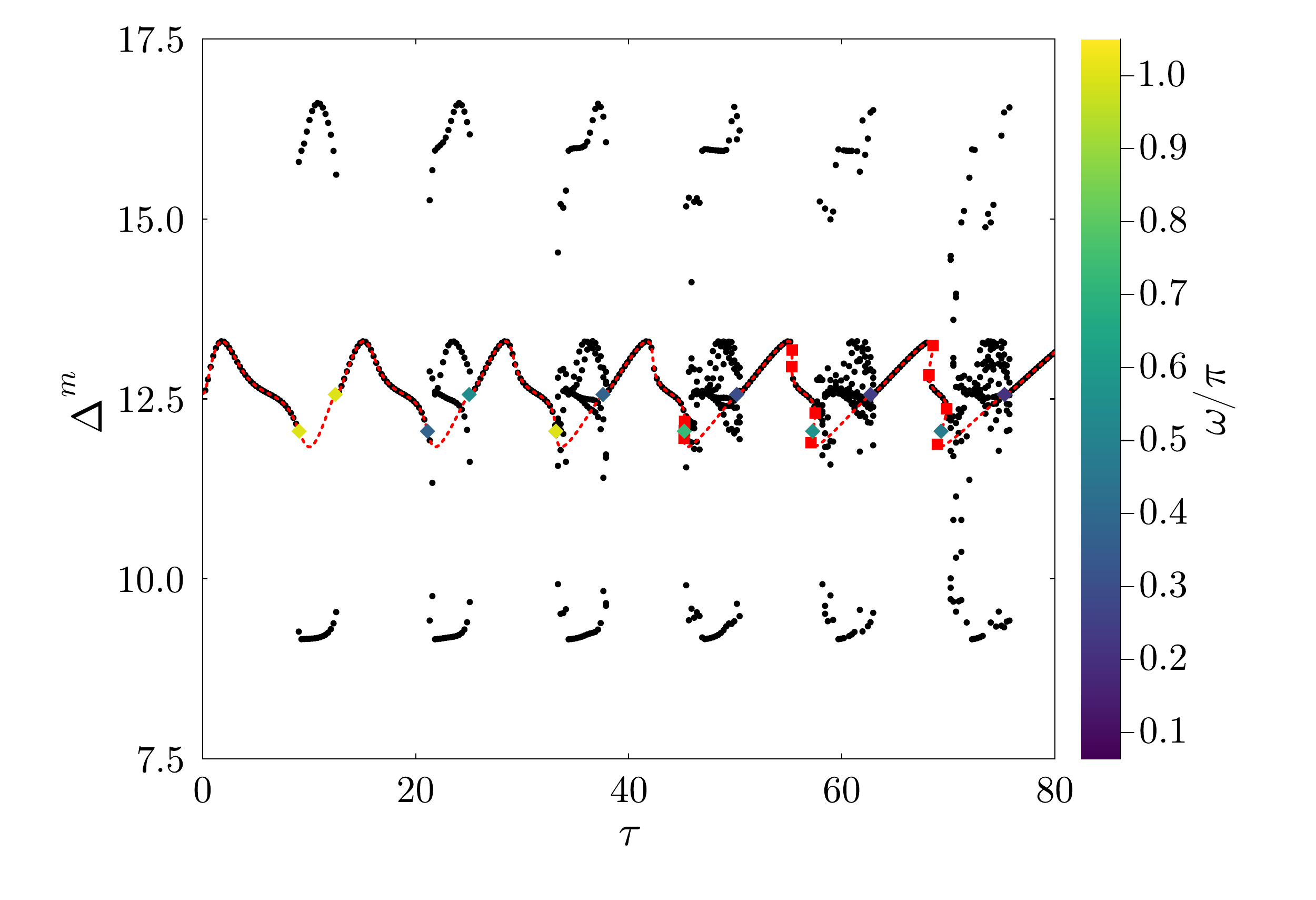}
\caption{Bifurcation structure of the Haken lighthouse autapse model in the $(\tau,\Delta)$ plane, combining semi-analytical branch information with direct numerical simulations. The dotted red curves show regular-spiking solution branches defined implicitly by $\mathcal{H}(T)=0$. Small red squares mark fold points obtained from direct solution of the double zero condition. Coloured diamond markers indicate dynamic-instability points obtained from direct solution of (\ref{dynamic}), with marker colour representing $\omega/\pi$. Black dots show late-time ISIs obtained from direct delayed simulations of the autapse model after discarding transients. Parameter values are $w_{0}=1$, $w_{\tau}=-1$, $\alpha=1$, $\beta=10$, $h=0$.
\label{Fig:Autapse}
}
\end{figure}

The resulting bifurcation structure is shown in Fig.~\ref{Fig:Autapse}.  The red dotted curves are the regular-spiking branches obtained by solving $\mathcal{H}(T)=0$.  Folds of these branches agree with the double-zero condition $\mathcal{E}'(0)=0$, shown by red square markers.  Dynamic instabilities obtained from (\ref{dynamic}) are shown by coloured diamond markers, with the colour encoding the modulation frequency $\omega/\pi$.  Direct numerical simulations of the delayed autapse model are superimposed as black points, showing the late-time ISIs after transients have been discarded.  Along stable portions of the regular-spiking branches the simulations collapse onto a single ISI value.  Beyond dynamic-instability points, the single ISI splits into multiple values, indicating jittering or higher-period ISI patterns.  Thus the figure illustrates both the multistability of regular spike trains and the loss of regular spiking through dynamic modulation of the ISIs.

\section{Phase-locked states in a reciprocal two-node network\label{Sec:N=2}}

We next consider a reciprocal two-node network.  This is the smallest network in which one can distinguish synchronous, anti-synchronous and asymmetric phase-locked states.  It therefore provides a useful bridge between the autapse calculation and larger networks with non-trivial phase structure.  We take $N=2$, with symmetric coupling and a common delay, so that $w_{ij}=\delta_{ij}+(1-\delta_{ij})w$ and $\tau_{ij}=(1-\delta_{ij})\tau$.  Introducing the phase difference $\phi=\phi_2-\phi_1$, the phase-locking equations (\ref{Phaselocked}) reduces to a pair of equations for $(T,\phi)$ given by $\mathcal{H}(T,\pm \phi)=0$.  Here $\mathcal{H}$ is $1$-periodic in its second argument and given by
\begin{equation}
\mathcal{H}(T,\phi) = \int_0^T \d t  \, S \left (P(t) + w P(t + \phi T -\tau) \right ) -2 \pi.
\label{H}
\end{equation}
From the form of (\ref{H}) we see that if there is a solution for a given value of $\tau$ the same solution are generated under the replacement $\tau \rightarrow \tau + kT$ for $k \in \ZSet$.
We note that a synchronous solution with $\phi=0$ satisfies $\mathcal{H}(T,0)=0$ and an anti-synchronous solution satisfies $\mathcal{H}(T,1/2)=0$, namely each solution is specified by only one equation that we write in the form $\mathcal{H}(T,\psi)=0$ for $\psi \in \{0,1/2 \}$.  For these cases we may construct the matrix $w(\lambda)$ as
\begin{equation}
w_{ij}(\lambda) = \frac{1}{T} \sum_{n\in \ZSet} f(\widetilde{\omega}_n (\lambda)) \,
\widehat{\eta} (\widetilde{\omega}_n (\lambda)) B_{ij} (\widetilde{\omega}_n (\lambda))  , 
\end{equation}
where 
\begin{equation}
B(\omega) = \begin{bmatrix}
1 & w \e^{\i \omega[\psi T-\tau]} \\
w \e^{\i \omega[\psi T-\tau]} & 1
\end{bmatrix} ,
\label{B}
\end{equation}
and
\begin{equation}
f(\omega) = \int_0^T \d t \, S' \left ( P(t) + w P(t+ \psi T - \tau) \right ) (\i \omega)\e^{\i \omega t} .
\end{equation}
The matrix $B$ has eigenvalues $\gamma_\pm(\omega) = 1 \pm w e^{\i \omega[\psi T-\tau]}$ whose eigenvectors $(1,1)$ and $(1,-1)$ are independent of $\omega$.  Thus we may diagonalise the linearised system (\ref{eigsystem}) by considering perturbations $\delta T$ parallel to these directions.  The synchronous and anti-synchronous solution are stable if $\text{Re} \, \lambda <0$ where $\mathcal{E}_\pm(\lambda) = 0$ and
\begin{equation}
\mathcal{E}_\pm(\lambda) = \left ( \e^\lambda-1 \right ) \dot{\theta} + \kappa - \Gamma_{\pm}(\lambda) .
\end{equation}
Here $\dot{\theta} = S(P(0) + w P(\psi T - \tau) )$ and
\begin{align}
\Gamma_{\pm}(\lambda)  & = \frac{1}{T} \sum_{n\in \ZSet} f(\widetilde{\omega}_n (\lambda)) \,
\widehat{\eta} (\widetilde{\omega}_n (\lambda)) \gamma_\pm(\widetilde{\omega}_n (\lambda)) ,\\
\kappa &= \frac{1}{T} \sum_{n \in \ZSet} f(\omega_n ) \,
\widehat{\eta} (\omega_n) \left [ 1+ w \e^{\i \omega_n [\psi T-\tau]}\right ] .
\end{align}
We now restrict attention to $\lambda \in \RSet$.  In this case $\mathcal{E}_\pm (\lambda) \in \RSet$ and we note the following:  $\mathcal{E}_+ (0) = 0$, $\mathcal{E}_- (0) = 2 R$,
where
\begin{align}
&R = \frac{w}{T} \sum_{n \in \ZSet} f(\omega_n ) \, \widehat{\eta} (\omega_n) \e^{\i \omega_n (\psi T-\tau)} \nonumber \\
&= w \int_0^T \d t \, S'(P(t) + w P(t+\psi T - \tau) ) P'(t +\psi T -\tau) \nonumber \\
& = \frac{1}{T} \PD{}{\psi} \mathcal{H}(T,\psi) .
\end{align}
Assuming that $f$ and $\widehat{\eta}$ are decaying functions of their arguments then we have that, for $T \leq  2 \tau$, $\kappa \geq \Gamma_{\pm}(\lambda)$ so that $\lim_{\lambda \rightarrow \infty} \mathcal{E}_{\pm}(\lambda) = \infty$.  
For $R <0$ the graph of $\mathcal{E}_-(\lambda)$ must cross the zero axis at a positive value of $\lambda$.  Hence, a solution (synchronous or anti-synchronous) is linearly unstable if $R <0$.  A double zero occurs at $R=0$ indicating a bifurcation at this point.
It is also possible to establish that changes in stability along solution branches, $T = T(\tau)$, occur at stationary points.
To see this, note that differentiation of $\mathcal{H}(T,\psi)=0$ with respect to $\tau$ gives
\begin{equation}
 \FD{T}{\tau} = \frac{RT}{\dot{\theta} + \psi RT}.
\end{equation}
Hence a bifurcation defined by $R=0$ coincides with a turning point of the curve $T=T(\tau)$, provided the denominator is non-zero.  In this way the branch geometry and the real-eigenvalue stability condition are directly linked.

For the case of phase-locked states that are neither synchronous or anti-synchronous we note that if $T=2 \tau$ 
then $\mathcal{H}(2 \tau, \phi) = 0 = \mathcal{H}(2 \tau, 1-\phi)$.  These equations are symmetric about $\phi=1/2$.  Hence, we need only solve the equation $\mathcal{H}(2 \tau, \overline{\phi}) = 0$ for $\overline{\phi} \in (0,1/2)$ and generate the solution for $\phi \in (1/2,1)$ from $1-\overline{\phi}$.
From the periodicity of the system we can generate related branches in the $(\tau,T)$ plane according to $((p+1/2)T,T)$, $p=1,2,3,\ldots$, for $T$ in the range of the symmetric branches.
We also note that if $\tau=T$ then there is a single equation for the phase given as the implicit solution of $\mathcal{H}(\tau, \overline{\phi})=0$, for $\overline{\phi} \in (0, 1/2)$ with the solution for $\phi \in (1/2,1)$ generated as $1-\overline{\phi}$.  
Other (asymmetric branch) solutions in the $(\tau,T)$ plane can be generated parametrically according to $(pT,T)$.  The stability of the asymmetric branches can be determined using the machinery developed around equation (\ref{M}).	
We expect asymmetric branches to connect maxima and minima of symmetric branches (where stability changes) in a pitchfork bifurcation. A proof is given in Appendix \ref{Appendix:Branches}.

\begin{figure}[htbp]
\centering
\includegraphics[width=\columnwidth]{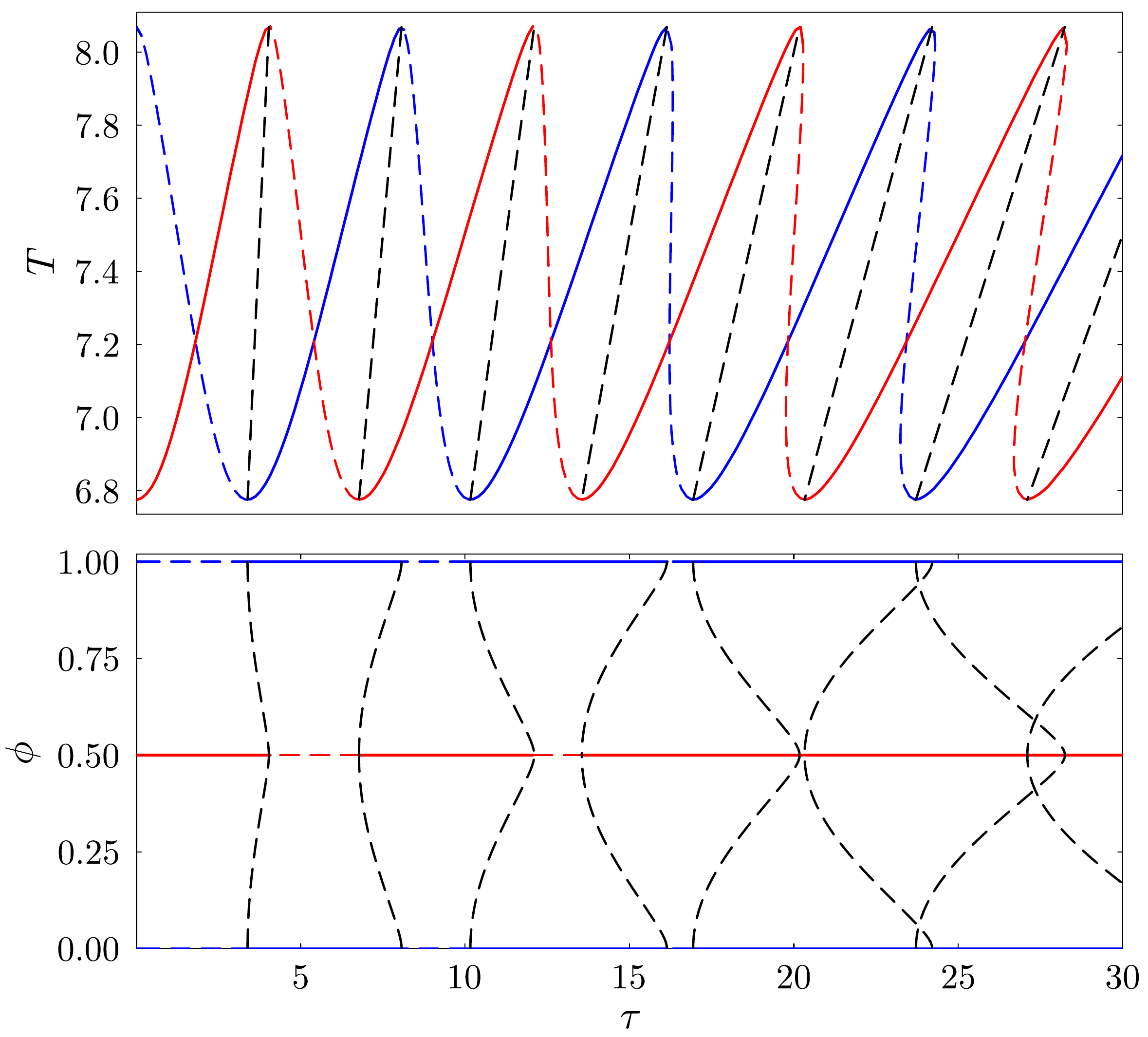}
\caption{Bifurcation diagram for a reciprocal two-node Haken network with a common delay $\tau$.  The upper panel shows the period $T=T(\tau)$, and the lower panel shows the corresponding phase difference $\phi=\phi(\tau)$.  Solid curves indicate stable phase-locked states and dashed curves indicate unstable states.  The branches with $\phi=0$ modulo $1$ correspond to synchrony, while branches with $\phi=1/2$ correspond to anti-synchrony.  Additional asymmetric branches occur in phase-conjugate pairs related by $\phi\mapsto1-\phi$ and attach to extrema of the symmetric branches.  Parameters are $\alpha=1$, $w=1$, $\beta=10$ and $h=0$.
\label{Fig:N=2Biftau}
}
\end{figure}

The resulting bifurcation structure is shown in Fig.~\ref{Fig:N=2Biftau}.  The upper panel shows the period $T$ as a function of the common delay $\tau$, while the lower panel shows the corresponding phase difference $\phi$.  Solid curves denote stable phase-locked states and dashed curves denote unstable states.  The synchronous branches have $\phi=0$ modulo $1$, whereas the anti-synchronous branches have $\phi=1/2$.  The asymmetric branches appear in pairs related by the transformation $\phi\mapsto1-\phi$ and connect neighbouring extrema of the symmetric branches.  Thus even in the reciprocal two-node network the delayed Lighthouse model supports multistability between synchronous and anti-synchronous phase-locked states (organised around unstable asymmetric ones).

\subsection{From two delays to one}

For a two node network we might also consider a situation with two distinct delays, with $\tau_{12} = \tau_1$ and $\tau_{21} = \tau_2$.  In this case the defining equations for a phase-locked state
are $\mathcal{H}(T,\phi;\tau_1) = 0$ and $\mathcal{H}(T,-\phi;\tau_2) = 0$, where $\mathcal{H}(T,\phi;\tau)$ is given by (\ref{H}).  Noting that the arguments to the integrands are $\phi T -\tau_1$ and $-\phi T-\tau_2$ it is natural to consider a new notion of phase and delay according to $\phi T -\tau_1 = \chi T - \overline{\tau}$ and $-\phi T -\tau_2 = -\chi T - \overline{\tau}$.  Solving this pair of equations gives $\overline{\tau} = (\tau_1+\tau_2)/2$ and $\chi = \phi-\delta/T$ where $\delta = (\tau_1-\tau_2)/2$.  We may then write $\mathcal{H}(T,\phi;\tau_1) = \mathcal{H}(T,\chi;\overline{\tau})$ and $\mathcal{H}(T,-\phi;\tau_2) = \mathcal{H}(T,-\chi;\overline{\tau}) $.   Thus, we are led to an equivalent problem with only a single effective mean delay that we must solve for $(\chi,T) = (\chi(\overline{\tau}),T(\overline{\tau}))$.  This can be done using the prescription above (for a single delay), from which we may construct $(\tau_1, \tau_2, T, \phi)$ parametrically in terms of $(\overline{\tau},\delta)$ as $(\overline{\tau} + \delta, \overline{\tau} - \delta, T(\overline{\tau}), \chi(\overline{\tau}) + \delta/T(\overline{\tau}) \! \mod 1)$.  
Consequently, unequal reciprocal delays do not introduce a new existence problem for phase-locked states in the two-node network: they shift the observed phase difference relative to the effective common-delay solution.  The case of two delays is taken up again later in Sec.~\ref{Sec:Numerics}.

\section{Phase-locked states in a ring with space dependent delays\label{Sec:Ring}}

Here we consider a Lighthouse network with interactions whose strength and delay both depend on distance.  For simplicity we consider nodes arranged on a ring with the distance between nodes $i$ and $j$ given by  $\operatorname{dist} (i,j) = \min(|i-j|, N-|i-j|)d$, for some spatial scale set by $d$.  The space-time connectome then has a circulant matrix representation with $w_{ij} = w_{\operatorname{dist} (i,j)}$ and $\tau_{ij} = \tau_{\operatorname{dist} (i,j)}$ with rows
generated by $w_{n} = w(\operatorname{dist} (0,n))$ for $n=0, \ldots,N-1$ for some function $w$ and 
delays chosen as $\tau_n = \operatorname{dist} (0,n)/v$.  Here $v>0$ represents a common conduction speed.
From the form of (\ref{Phaselocked}) we can expect the emergence of \textit{twisted} states defined by $\phi_i = i q /N$, so that $\phi_i - \phi_j = (i-j)q/N$ $(\! \! \! \! \mod 1)$ for some $q \in \ZSet$.  For $q=0$ this would represent the synchronous state and for $q=1$ the so-called splay state.  
We note that mode $q$ and $N-q$ are conjugate (reversed) twisted states.
For a fixed value of $q$ the emergent period $T$ is given implicitly from (\ref{Phaselocked}) as the solution to $\mathcal{H}_q(T)=0$, where
\begin{equation}
\mathcal{H}_q(T) = \int_{0}^{T} \d t \, S \left ( \sum_{n=0}^{N-1} w_{n} P(t + A_{n,q}) \right ) -2 \pi ,
\end{equation}
and $A_{n,q} = n q T/N - \tau_{n}$ with $w_n=w_{N-n}$ and $\tau_n=\tau_{N-n}$ (reflecting the fact that both coupling strength and delay depend only on ring distance).
The circulant properties of the space-time connectome mean that the matrix $\mathcal{L}(\lambda)$ is also circulant.  Specifically, the matrix $G(\lambda)$ has a row generator given by (\ref{Gl}) under the replacement $ij \rightarrow n$ and $f_i (\omega) \rightarrow f(\omega) = \int_0^T \d t \, S'(X_q(t)) (\i \omega) \e^{\i \omega t}$ for $n=0,\ldots,N-1$ and $X_q(t) = \sum_{n=0}^{N-1} w_n P(t+A_{n,q})$.  Similarly we will denote the row generators for $w_{ij}(\lambda)$ using the notation $w_n(\lambda)$.

In general for $\lambda \neq 0$ the matrices $w(\lambda)$ and $w(0)$ defined after equation (\ref{eigsystem}) will not have a common eigenspace.  However, when these are circulant they do and we may exploit this to diagonalise (\ref{eigsystem}).    After introducing right (left) normalised eigenvectors of $w(0)$ as $r^\mu$ ($l^\mu$), we can write $w(\lambda) = \sum_{\mu} \gamma_\mu(\lambda) l^\mu \otimes r^\mu$, where $\otimes$ denotes the tensor product.  Here,  $\gamma_\mu(\lambda)$ can be constructed by projection of $w(\lambda)$ on to $l^\mu$ and $r^\mu$ as
$\gamma_\mu(\lambda) = \left ( l^\mu \right )^\top w(\lambda) r^{\mu} =  \sum_{i,j} l^\mu_i w_{ij}(\lambda) r^\mu_j$.  
By introducing the row-sum $\kappa = \sum_{n=0}^{N-1} w_{n}(0)$ and considering the vector $\delta T$ to be parallel to a right eigenvector of $w(0)$ we can diagonalise (\ref{eigsystem}) so that the matrix in equation (\ref{M}) becomes $\left [ \left (\e^{\lambda} - 1 \right ) \dot{\theta}  + \kappa -\gamma_\mu(\lambda) \right ] I_N$,
where $\dot{\theta} = S(\sum_{n=0}^{N-1} w_n P(A_{n,q}))$ and $I_N$ is the $N \times N$ identity matrix.
For non-trivial solutions we must have that $\mathcal{E}_\mu(\lambda) = 0$, where 
\begin{equation}
\mathcal{E}_\mu(\lambda) = \left (\e^{\lambda} - 1 \right ) \dot{\theta}  + \kappa -\gamma_\mu(\lambda) , \qquad \mu =1, \ldots, N.
\label{E}
\end{equation}
Since $w(0)$ is symmetric the left and right eigenvectors are related by a transpose and we have explicitly that
\begin{equation}
\gamma_\mu(\lambda) = \sum_{n=0}^{N-1} w_n(\lambda) \exp \left (
\frac{2 \pi {\rm i} (\mu-1) n}{N}
\right ) ,
\end{equation}
The eigenvector associated with eigenvalue $\gamma_\mu$ is given by
$l^\mu = (1,a_\mu, a_\mu^2, \ldots, a_\mu^{N-1})/\sqrt{N}$, where $\mu=1,\ldots, N$, and $a_\mu=\exp(2 \pi \i (\mu -1)/N)$ are the $N$th roots of unity. 
In this way we may determine the stability of a phase-locked state according to the condition $\text{Re}\, (\lambda) <0 $ for all $\mu$ (excluding the zero eigenvalue that arises from phase-shift symmetry).

\begin{figure*}[htbp]
\centering
\includegraphics[width=0.75\textwidth]{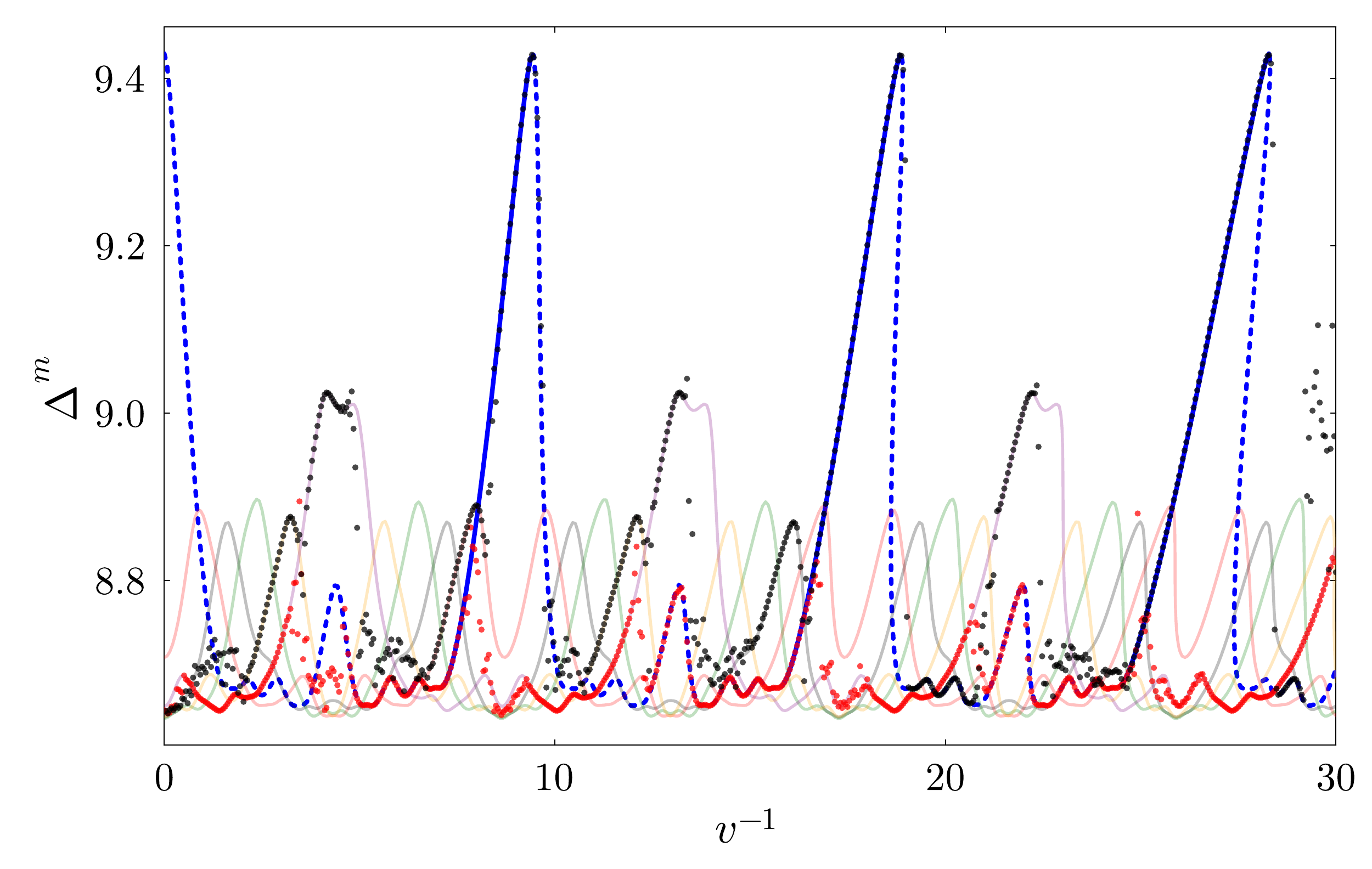}
\caption{Branch structure and quasi-static sweep simulation for an $11$-node ring Haken network with distance-dependent delays.  The branch curves show $\Delta=T$ as a function of $v^{-1}$ for $w(x)=\exp(-|x|/\sigma)/(2\sigma)$ and primary twisted states $q=0,1,\ldots,5$.  The synchronous branch $q=0$ is shown in blue, with solid segments indicating stability with respect to the real-mode diagnostic and dashed segments indicating real-mode instability.  The remaining branches are shown as fixed-delay organising branches: splay, $q=1$ in red, and twisted states $q=2,3,4,5$ in black, green, orange and purple, respectively.  Direct delayed simulations are overlaid by slowly sweeping $v^{-1}$ from $30$ to $0$ and then back from $0$ to $30$, recording late instantaneous ISIs in small bins of $v^{-1}$.  Red points indicate the downward sweep from high to low $v^{-1}$, while black points indicate the upward sweep from low to high $v^{-1}$.  The sweep reveals direction-dependent tracking of attracting branch segments and anticipates the adaptive conduction-speed dynamics considered in Sec.~7.  Parameters are $\beta=10$, $h=0$, $\alpha=1$, $d=1$ and $\sigma=3$.\label{Fig:N=11Bif}}
\end{figure*}

If we restrict attention to $\lambda \in \RSet$ this does not mean that $\mathcal{E}_\mu(\lambda)  \in \RSet$ since the Fourier modes occur in complex-conjugate pairs, and the modal characteristic functions are complex-valued even for real $\lambda$.  However, for the synchronous state ($q=0$), the modal characteristic functions are real on the real axis and real-root instabilities can be detected by the sign of $\mathcal{E}_\mu(0)$.
To see this note that $\lim_{\lambda \rightarrow \infty} \mathcal{E}_\mu(\lambda) >0$ and so if $\mathcal{E}_\mu(0) < 0$ there must be a positive real root, and hence an instability.
Thus for the synchronous case with $\gamma_\mu(0) \in \RSet$, a real instability occurs when $\mathcal{E}_\mu(0) = \kappa-\gamma_\mu(0)=0$ for some $\mu\neq 1$ with
\begin{equation}
\mathcal{E}_\mu(0) = \sum_{n=0}^{N-1} a_n \left [1 - \cos \left ( \frac{2 \pi (\mu-1) n}{N} \right) \right ],
\end{equation}
where $a_n=a_{N-n}$ and $a_n=w_n \int_0^T S'(X_0(t)) P'(t+A_{n,0}) \d t = w_n  \sum_m Y_{-m},(\i\omega_m) \widehat{\eta}(\omega_m)\,\e^{-\i\omega_m\tau_m}$ for $n=0,\ldots,N-1$, where $Y_m$ are the Fourier series coefficients of $S'(X_0(t))$.


For a nonzero twisted state, the twist introduces an orientation. Even though $w_n=w_{N-n}$ and $\tau_n=\tau_{N-n}$,  the coefficients contain the signed phase shift
$A_{n,q}=nqT/N-\tau_n$, so generally $a_n\neq a_{N-n}$.  Hence, the circulant matrix (in the linear stability analysis) is real but not symmetric. Hence its Fourier eigenvalues come in complex-conjugate pairs, and
$\mathcal{E}_\mu(\lambda) \in \CSet$ for real $\lambda$.  In these cases we must use the more general prescription for determining stability as described above.

Figure~\ref{Fig:N=11Bif} illustrates the resulting branch structure for an $11$-node ring.  The parameter plotted on the horizontal axis is $v^{-1}$, so moving to the right corresponds to slower conduction and hence larger distance-dependent delays.  The branch curves are obtained by solving $\mathcal{H}_q(T)=0$ for the primary twisted states $q=0,1,\ldots,5$.  The synchronous branch is plotted with a stability diagnostic based on the real-mode condition above, using solid line segments where no real instability is detected and dashed line segments where a real instability is present.  The remaining twisted branches are shown as organising branches for the fixed-delay problem.

The same figure also overlays direct simulations in which $v^{-1}$ is slowly swept through the range of the bifurcation diagram.  The network is first allowed to relax at the high-delay end, $v^{-1}=30$, after which $v^{-1}$ is swept down from $30$ to $0$ and then swept back up from $0$ to $30$.  At each stage the instantaneous ISIs are recorded, retaining the late ISI values in small bins of $v^{-1}$ to produce a quasi-static sweep trace.  Red points correspond to the downward sweep from high to low $v^{-1}$, while black points correspond to the upward sweep from low to high $v^{-1}$.  The comparison between swept simulations and frozen-delay branches shows how the network follows different attracting parts of the branch structure depending on sweep direction, with the possibility of hysteresis and branch switching.  This use of a prescribed sweep is not itself an adaptive-delay model, but it anticipates the next section, where conduction speeds and delays evolve dynamically through activity-dependent white matter plasticity.

\section{White matter plasticity\label{Sec:WMPlasticity}}

Myelin is a fatty substance that wraps and electrically insulates axons and gives them a white colour.  This \textit{white matter} makes up roughly 50\% of the human brain and is vitally important for controlling communication pathways via the modulation of axonal conduction speeds.
The temporal synchrony of signals arriving from different neurons or brain regions is known to be essential for proper neural processing \cite{Gansel2022}.  Nonetheless, how this is achieved and maintained in delayed, complex networks remains incompletely understood \cite{Pajevic2014}.  
Experimental results show that oligodendrocytes (a type of glial cell that produces myelin) are responsive to changes in neural activity and that myelination is activity dependent. Optogenetic and electrical manipulations have shown that increases in firing activity promote myelination. Fluctuations in neural activity engage oligodendroglia, resulting in changes in axonal conduction speeds through the formation of new myelin segments, adaptive changes in myelin sheath thickness and/or internodal length \cite{Mount2017,Gibson2014,Xin2020}. While neuron-oligodendrocyte signalling remains poorly understood and a topic of intense research, experiments indicate that neural firing rates and axonal conduction speeds are generally positively correlated \cite{Noori2020}.  Thus, there is now a consensus that white matter is \textit{plastic}, namely that the myelination of an axon is activity dependent  see e.g.,  \cite{Fields2008,Gibson2014,Vivo2019}.  Given that myelination affects conduction speed, the delays that we have considered as fixed up until now may themselves evolve in a state-dependent fashion.
The time-scale for white matter plasticity can be as short as days to weeks in the context of a new learning experience, though ranges up to years \cite{Huber2018}.  This has recently begun to be explored from a modelling perspective by Lefebvre and colleagues \cite{Noori2020,Park2020,Talidou2021,Talidou2022,Lefebvre2025}, with recent work by Fields and colleagues emphasising the role of oligodendrocyte-mediated myelin plasticity in facilitating neural synchronisation, whereby white matter plasticity may help distant brain regions remain synchronised by compensating for long transmission times through increased conduction speeds \cite{Pajevic2014,Pajevic2023}. 

Building on the work of Lefebvre and colleagues it is natural to describe white matter plasticity in terms of a biologically motivated rule for the evolution of conduction speed.  For a fixed axonal fibre length between nodes $i$ and $j$ of length $d_{ij}$, we will consider a uniformly myelinated system that gives rise to a delay $\tau_{ij}(t) = d_{ij}/c_{ij}(t)$ for a signal travelling along the fibre with speed $c_{ij}$.  In the absence of any myelination we will define this speed to be $c^0_{ij}$, with a corresponding fixed delay $\tau_{ij}^0 = d_{ij}/c_{ij}^0$, such that with an increase in myelination $c_{ij} > c_{ij}^0$.  To develop a plasticity rule it is important to bear in mind that i) myelination is activity dependent (via axon-glia interaction), and ii) axonal plasticity is slow and has a metabolic cost.  Within the context of the spiking Haken Lighthouse model presented here we propose a phenomenological model of the form
\begin{equation}
\tau_w \FD{}{t} c_{ij} = c_{ij}^0 - c_{ij} + f_{ij} (a_{ij}), 
\label{WM1}
\end{equation}
with 
\begin{align}
\label{WM2}
a_{ij} (t) &= \int_{t-\tau_{ij}(t)}^t \d t' \, S \left ( \psi_j(t') \right ) ,\\
\label{WM3}
f_{ij}(a) &= \kappa_{ij}\tanh(\sigma_{ij} (a-h_{ij})) .
\end{align}
In this model, changes in conduction speed are governed by the interplay between two general mechanisms, namely myelin \textit{retraction} and myelin  \textit{formation}. 
Equation (\ref{WM2}) has an integral over the time that it takes for a signal to propagate along a fibre.  This is meant to help capture the fact that metabolic cost will depend upon activity over the whole length of the fibre whilst it is active.  Here, we take the firing rate $S(\psi_j)$ as a natural proxy for the activity of node $j$ that sends signals to node $i$.  If this sustained activity is high with respect to some threshold then we would expect conduction speeds to increase, whilst if it is low then this would lead to a decrease.  We model this nonlinear behaviour with the use of the function $f_{ij}$ via a simple $\tanh$ shape.   Note that on each edge $(i,j)$ we have parameters for baseline speed $c_{ij}^0$, threshold $h_{ij}$, steepness of the saturating nonlinearity $\sigma_{ij}$, and strength of modulation $\kappa_{ij}$.  Finally, $\tau_w$ sets the slow time-scale of white matter myelination relative to fast neuronal oscillations.

A useful form for $a_{ij}$ that can be readily implemented numerically is obtained by differentiation as
\begin{equation}
\FD{}{t} a_{ij} (t) = S \left ( \psi_j(t) \right ) - S \left ( \psi_j(t-\tau_{ij}(t)) \right ) \left [ 1- \FD{}{t} \tau_{ij} (t) \right ]
\end{equation}
with initial data $a_{ij} (0) = \int_{-\tau_{ij}(0)}^0 \d t \, S \left ( \psi_j(t) \right )$ (if the history is specified for $t\leq 0$).
In contrast to previous white matter plasticity rules for phase-oscillator networks as in \cite{Karimian2019,Park2020} the evolution for speed is governed by a state-dependent delay differential equation as opposed to an ordinary differential equation (ODE).

The next subsection exploits the time-scale separation between the fast spiking dynamics and the slow conduction-speed dynamics to interpret this adaptive system from a slow--fast perspective.

\subsection{A \textit{slow--fast} perspective\label{Sec:perspective}}

The adaptive system combines fast spiking dynamics with slow changes in conduction speed.  It is therefore natural to view it as a slow--fast problem, with the static-delay theory of the previous sections providing the frozen fast subsystem.  
Writing $X$ for the collection of fast Lighthouse variables $(\theta_i, \psi_i, \dot{\psi}_i)$ and $c$ for the matrix of conduction speeds ($[c]_{ij}=c_{ij}$), the full adaptive system takes the abstract form
\begin{equation}
\dot X = F(X;c),
\qquad
\tau_w \dot c = G(X,c),
\label{eq:fastslowabstract}
\end{equation}
where the delays entering the fast subsystem are slaved to the conduction variables through $\tau_{ij}=d_{ij}/c_{ij}$. Introducing $\varepsilon=\tau_w^{-1}\ll 1$ gives the singularly perturbed form
\begin{equation}
\dot X = F(X;c),
\qquad
\dot c = \varepsilon G(X,c).
\label{eq:epsform}
\end{equation}
Thus the spiking dynamics evolves on the fast timescale while the conduction variables drift through delay space on the slow timescale.
It is worth stressing that the fast subsystem is not a typical smooth ODE flow. Rather, it is naturally formulated in terms of event times and spike trains (and the linear stability theory developed earlier is correspondingly written in terms of perturbations of firing times). This event-driven structure is one of the attractive features of the Lighthouse model, but it also means that some of the standard hypotheses and proofs from geometric singular perturbation theory for smooth finite-dimensional flows cannot be imported wholesale \cite{Jones1995}. In particular, when speaking of critical manifolds, normal hyperbolicity, and slow manifolds below, one should understand these statements at a formal or geometric level, guided by the spectral stability theory of the frozen event-driven problem. A fully rigorous invariant-manifold theory for the adaptive event-driven system would require additional work, especially when recognising that the slow flow is governed by a state-dependent delay differential equation.

For $\varepsilon=0$ the conduction variables are frozen, and the fast subsystem reduces to the static-delay Lighthouse problem analysed earlier (and parameterised by $c$). 
A branch of frozen phase-locked states therefore defines a critical manifold
\begin{equation}
\mathcal M_0=\{(X^*(\cdot;c),c): c \text{ lies on the branch}\},
\label{eq:criticalmanifold}
\end{equation}
in the extended fast--slow state space. Here $X^*(\cdot;c)$ denotes the corresponding frozen phase-locked orbit.
In this case the activity variable $a_{ij}$ depends on how the delay $\tau_{ij}$ compares with the period $T$. If
\begin{equation}
\tau_{ij}=(p_{ij}+r_{ij})T,
\qquad
p_{ij}\in\mathbb Z_{\ge 0},
\qquad
r_{ij}\in[0,1),
\label{eq:commensurate}
\end{equation}
then the integration window in \eqref{WM1} spans $p_{ij}$ full periods together with a fractional interval of length $r_{ij}T$, so that
\begin{equation}
a_{ij}(t)=2\pi p_{ij}+\int_{t-r_{ij}T}^{t} \d t'\, S\big(\psi_j(t')\big).
\label{eq:aijsplit}
\end{equation}
An especially interesting case is the integer family of commensurate states (e.g., the asymmetric branches of Sec.~\ref{Sec:N=2}):
\begin{equation}
\tau_{ij}=p_{ij}T,
\qquad
p_{ij}\in\mathbb Z_{\ge 0},
\label{eq:integerfamily}
\end{equation}
for which the residual term in \eqref{eq:aijsplit} vanishes and hence
\begin{equation}
a_{ij}(t)=2\pi p_{ij}.
\label{eq:aijinteger}
\end{equation}
Thus, the slow adaptive law closes exactly on the corresponding phase-locked branch to yield the autonomous ODE:
\begin{equation}
\tau_w \dot c_{ij} = c_{ij}^0-c_{ij}+\kappa_{ij}\tanh\!\big(\sigma_{ij}(2\pi p_{ij}-h_{ij})\big).
\label{eq:exactslowflow}
\end{equation}
Thus the nonlinear plasticity rule retains the special role of commensurate phase-locked states, but the adaptive forcing now depends on the signed mismatch $2\pi p_{ij}-h_{ij}$.

The relevant notion of normal hyperbolicity is inherited from the event-driven linear stability problem for the frozen phase-locked state. After quotienting out the neutral direction associated with global phase shift, a branch is normally attracting if every non-trivial fast exponent satisfies
$\text{Re}\,\lambda<0$, with $\lambda$ determined as $\mathcal{E}(\lambda)=0$ as described in Sec.~\ref{Sec:Phaselocked}.
When this holds, it is natural to regard $\mathcal M_0$ as a normally attracting critical manifold. Formally, and in the spirit of geometric singular perturbation theory, one then expects that for sufficiently small $\varepsilon$ there exists a nearby locally invariant slow manifold $\mathcal M_\varepsilon$ that perturbs smoothly from $\mathcal M_0$. Trajectories first relax rapidly toward $\mathcal M_\varepsilon$ in the fast directions and then drift slowly along it under the induced adaptive dynamics.

A commensurability condition such as \eqref{eq:integerfamily} is not by itself sufficient to define an adaptive steady state. One must also satisfy the kinematic relation $\tau_{ij} = d_{ij}/c_{ij}$ and the frozen phase-locking conditions that determine the relevant branch $T=T(c)$.
A fixed point $c^*$ of the reduced slow flow on an integer-commensurate branch therefore satisfies
\begin{equation}
c_{ij}^*=c_{ij}^0+\kappa_{ij}\tanh\!\big(\sigma_{ij}(2\pi p_{ij}-h_{ij})\big),
\label{eq:cstarinteger}
\end{equation}
together with the compatibility condition
\begin{equation}
\frac{d_{ij}}{c_{ij}^*}=p_{ij}T(c^*).
\label{eq:compatibility}
\end{equation}
Hence the existence of a slow fixed point on a $p_{ij}T$ family is a joint condition involving the frozen phase-locked branch, the anatomical distances $d_{ij}$, and the plasticity parameters $(c_{ij}^0,\kappa_{ij},\sigma_{ij},h_{ij})$.
This highlights that a candidate adaptive phase-locked state is selected not merely by the existence of a static phase-locked solution, but by a mutual compatibility between the fast branch geometry and the nonlinear plasticity rule. In particular, the thresholds $h_{ij}$ can be used to bias the slow dynamics toward some commensurability classes and away from others.

When \eqref{eq:integerfamily} does not hold exactly, the quantity $a_{ij}(t)$ retains an explicit dependence on the phase of the fast orbit through \eqref{eq:aijsplit}. 
On a normally attracting branch of frozen phase-locked states one may then pass to a reduced slow flow on the corresponding slow manifold.  This gives rise to a non-autonomous ODE that can be simplified to an autonomous form after realising that \textit{averaging} can be legitimately applied.
Averaging is suitable when there is a clear timescale separation, so that the fast spiking dynamics settles rapidly onto a frozen $T$-periodic phase-locked orbit while the conduction variables change only by $O(\varepsilon)$ over one fast cycle. In that setting, the slow variables effectively see only the cycle-averaged forcing from the fast subsystem with $a_{ij}(t) = a_{ij} (t+ T(c))$. 
If $X^*(t;c)$ denotes the corresponding frozen $T(c)$-periodic orbit, then the averaged reduced slow flow is
\begin{equation}
\tau_w \dot c_{ij}
=
c_{ij}^0-c_{ij}
+
\frac{\kappa_{ij}}{T(c)}\int_0^{T(c)}
\tanh\!\big(\sigma_{ij}(a_{ij}(t;c)-h_{ij})\big) \d t.
\label{eq:exactaveragedslowflowtanh}
\end{equation}
A slow fixed point $c^*$ on the branch therefore satisfies
\begin{equation}
c_{ij}^*=c_{ij}^0+
\frac{\kappa_{ij}}{T(c)}\int_0^{T(c)}
\tanh\!\big(\sigma_{ij}(a_{ij}(t;c^*)-h_{ij})\big) \d t.
\label{eq:generalfixedpoint}
\end{equation}
together with the requirement that $c^*$ lie on the frozen phase-locked branch under consideration.

Linearising the exact reduced slow flow \eqref{eq:exactaveragedslowflowtanh} at a reduced slow fixed point gives
\begin{equation}
\tau_w \dot \eta = \D F(c^*)\eta,
\label{eq:slowlin}
\end{equation}
where
\begin{align}
&[\D F(c)]_{(ij),(k\ell)}
=
-\delta_{ik}\delta_{j\ell} \nonumber \\
&+
\frac{1}{T(c)}\int_0^{T(c)}
 f'_{ij}\!\big(a_{ij}(t;c)\big)
 \frac{\partial a_{ij}}{\partial c_{k\ell}}(t;c)\d t,
\label{eq:DF}
\end{align}
and $f'_{ij}(a)=\kappa_{ij}\sigma_{ij}\operatorname{sech}^2\!\big(\sigma_{ij}(a-h_{ij})\big)$.
The factor $f'_{ij}(a)$
controls how strongly each edge responds to perturbations. Edges operating near threshold are highly plastic, whereas edges in a saturated regime are effectively pinned.
The full adaptive state is attracting when the frozen phase-locked orbit is spectrally stable in the fast subsystem and the reduced slow fixed point is stable within the slow manifold.  We note that for the commensurate states given by (\ref{eq:integerfamily}) 
the slow flow is given exactly by \eqref{eq:exactslowflow} and since
$\partial a_{ij} / \partial c_{k\ell}(t;c) = 0$ along the exact commensurate manifold where $a_{ij}=2\pi p_{ij}$ (so that  $\D F$ only has eigenvalues $-1$) steady states in the slow subsystem are stable (though may be unstable in the fast subsystem).


If a reduced slow fixed point is stable along the slow manifold but the frozen phase-locked orbit is unstable in a fast direction, then it can act as an organising saddle for the full dynamics. Trajectories may be drawn toward it in conduction space and linger near it for long times, but are ultimately expelled in the transverse fast directions. Such states naturally organise switching, bottlenecks, and delayed escape.
The reduced slow description is valid only while the underlying branch remains normally hyperbolic. When the frozen fast subsystem loses stability (for example at a fold, a symmetry-breaking point, or a dynamic instability)  normal hyperbolicity fails and the reduced single-manifold description breaks down. One then expects a combination of slow drift along the attracting portion of a branch and fast jumps to other branches. In this way the geometric singular perturbation viewpoint naturally accommodates convergence to a stable adaptive state, switching between competing phase-locked states, and delayed departure from weakly unstable states.
The nonlinear plasticity rule also suggests a natural mechanism for relaxation oscillations. Suppose there are two competing phase-locked branches, each normally attracting for the frozen fast subsystem over some interval, but such that the induced slow flow along one branch points toward larger delays while the induced slow flow along the other points toward smaller delays. Then the system may drift slowly along the first branch until normal hyperbolicity is lost, undergo a rapid jump to the second branch, drift slowly back in the opposite direction, and repeat. This produces a slow--fast alternation between branchwise drift and fast transitions, i.e. a relaxation oscillation in conduction (or delay) space.
By contrast, if the slow flow does \emph{not} reverse direction between nearby stable branches, then persistent switching is less likely. In that situation trajectories may instead approach a final slow steady state, either on an exactly commensurate branch satisfying \eqref{eq:cstarinteger}--\eqref{eq:compatibility} or on a nearby non-commensurate branch satisfying \eqref{eq:generalfixedpoint}. Thus the nonlinear rule supports both possibilities: sustained slow switching when branchwise drifts oppose one another, and convergence to a final adaptive state when the slow drift is globally directed toward a compatible attractor.

\subsection{Numerical illustrations\label{Sec:Numerics}}

Here we revisit the study from Sec.~\ref{Sec:N=2} with fixed delays and allow them to become plastic according the white matter plasticity rule introduced above.  With the introduction of the plasticity rule (\ref{WM1})-(\ref{WM3}) we use this to illustrate two key observations concerning i) the evolution to a stationary phase-locked state, and ii) the opportunity for oscillations that switch between states.
To make a comparison between theory and simulations it is useful to complement the notion of a firing event with a firing phase.  We can generate a phase associated with the firing times of each neuron as follows. Let $\mathcal{T}_i^m$ be the set of firing times lying in the interval $[T_i^m, T_i^{m+1})$. If there is only one spike in this interval, we transform it into a phase $\phi_i^m \in [0,1)$ using the transformation \cite{Pinsky1995}
\begin{equation}
\phi_i^m=\frac{\mathcal{T}_i^m-T_i^m}{T_i^{m+1} - T_i^m}.
\label{InstPhase}
\end{equation}

In the earlier reciprocal two-node problem presented in Sec.~\ref{Sec:N=2}, the frozen branch diagrams are computed on the equal-delay subspace $\tau_{12} = \tau_{21}= \tau$.  
We note that (for the reciprocal pair) the averaged slow flow given by (\ref{eq:exactaveragedslowflowtanh}) is symmetric under exchange of the two directed edges provided the edge parameters are matched and the frozen fast branch is itself symmetric (namely the synchronous and anti-synchronous solutions).  A similar statement holds for slow flow on the asymmetric commensurate branch as long as we impose the further condition that $p_{12}=p_{21}$.  For the purposes of using Fig.~\ref{Fig:N=2Biftau} as a backbone for organising the full dynamics with white matter plasticity we shall restrict ourselves to this subspace.

\begin{figure}[htbp]
\centering
\includegraphics[width=\columnwidth]{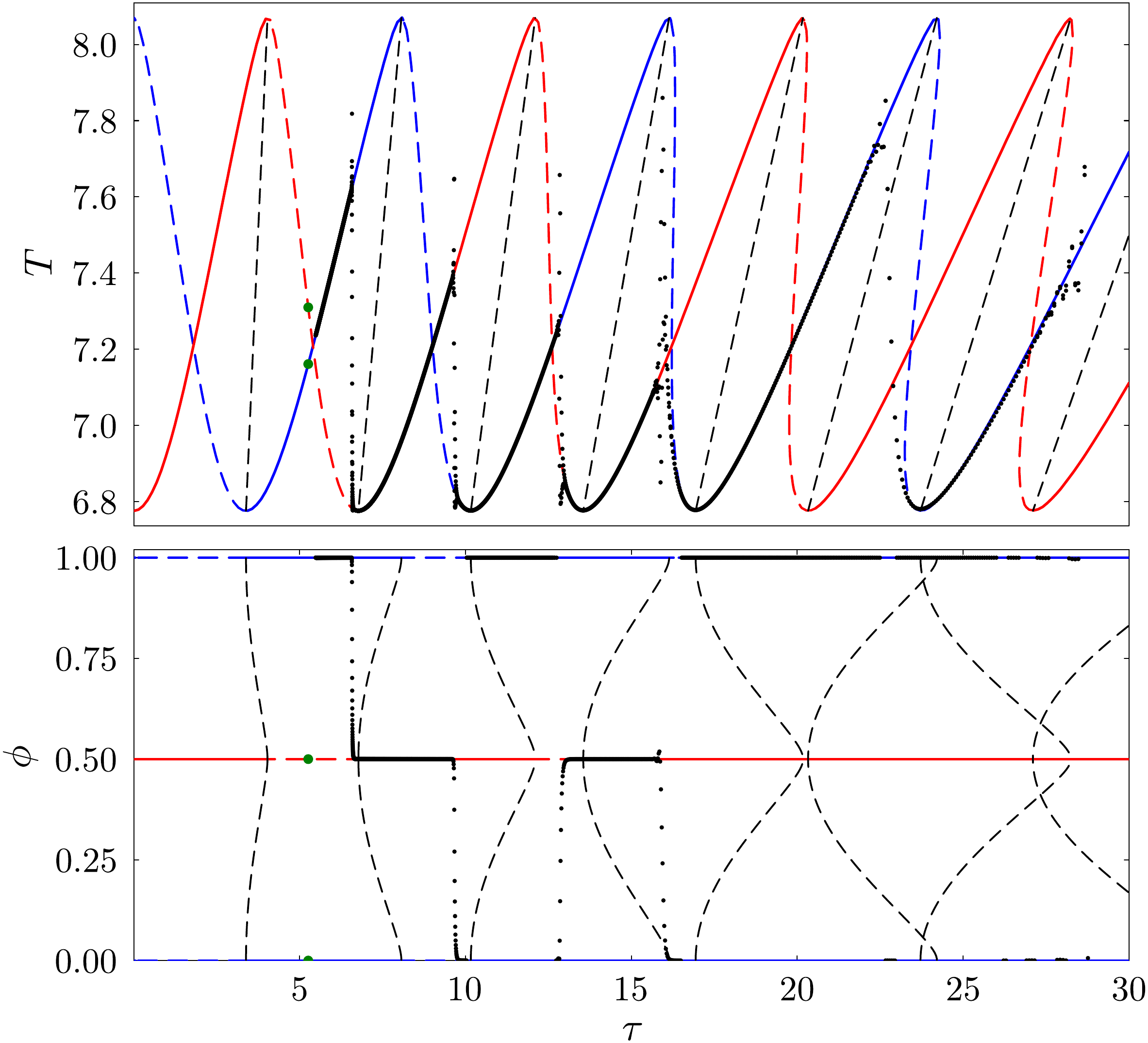}
\caption{Numerical simulation of the reciprocal two-node Haken model with white matter plasticity on the symmetric subspace $\tau_1=\tau=\tau_2$, where $\tau_1=\tau_{12}$ and $\tau_2=\tau_{21}$.  The upper panel shows instantaneous ISIs for both neurons against the current delay $\tau$ at firing events, and the lower panel shows the corresponding instantaneous phase defined by \eqref{InstPhase}.  The coloured curves are the frozen-delay branches from the two-node bifurcation diagram Fig.~\ref{Fig:N=2Biftau}, while black points show the adaptive simulation projected onto the same coordinates.  Filled green circles mark fixed points of the reduced slow flow on symmetric branches.  The simulation shows rapid attraction toward a phase-locked state, slow drift along the frozen branch structure, switching between stable symmetric branches, and final approach to a slow fixed point.  Parameters are as in Fig.~\ref{Fig:N=2Biftau}, with $d_{12}=1=d_{21}$, $\tau_{12}^0=5=\tau_{21}^0$, $\kappa_{12}=0.01=\kappa_{21}$, $h_{12}=8\pi=h_{21}$, $\sigma_{12}=1=\sigma_{21}$, and $\tau_w=10^4$.  The initial delay history is constant with $\tau_{12}(0)=30=\tau_{21}(0)$.
\label{Fig:N=2FP}
}
\end{figure}

\begin{figure}[htbp]
\centering
\includegraphics[width=\columnwidth]{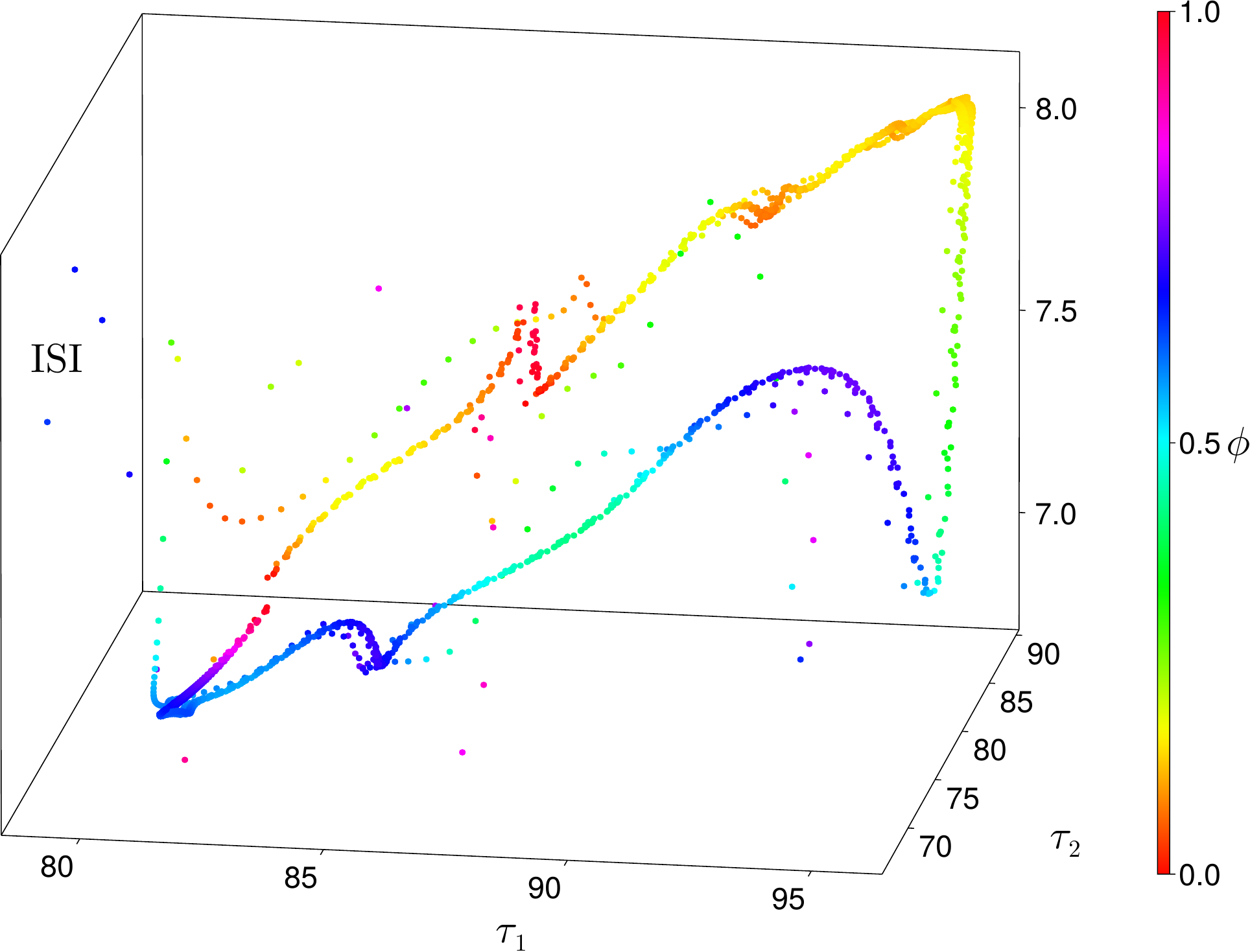}
\caption{Numerical simulation of the reciprocal two-node Haken model with two plastic delays showing $(\text{ISI},\tau_1,\tau_2)$.  The colour of data points indicates the corresponding instantaneous phase.
The simulations show switching between low and high ISIs.
Parameters are the same as in Fig.~\ref{Fig:N=2FP} with $(\kappa_{12},\kappa_{21}) = (0.1,0.1)$, $(\tau_{12}^0, \tau_{21}^0)  = (90,80)$, and $(h_{12},h_{21})=(74.4,69.8)$.
\label{Fig:N=2Relaxation}
}
\end{figure}

In Fig.~\ref{Fig:N=2FP} we show a simulation of the full adaptive system on the symmetric subspace $\tau_1 = \tau = \tau_2$.  Here we plot the instantaneous ISI (for both neurons) and the corresponding value of $\tau$ at a firing event in the upper panel and a similar plot using instantaneous phase in the lower panel.  
The coloured curves show the frozen-delay branches from the two-node bifurcation diagram, while the black simulation points show the trajectory of the adaptive system projected onto this frozen branch structure.  The trajectory first relaxes rapidly toward a phase-locked state and then evolves slowly as the conduction delay changes.  During this slow drift it switches between stable symmetric branches before approaching a fixed point of the reduced slow flow, indicated by a filled green marker.  This behaviour is consistent with the slow--fast picture: the adaptive dynamics is organised by attracting pieces of frozen phase-locked branches, while the slow plasticity rule moves the system through delay space.
Given that multi-stability of solutions increases with higher values of a frozen $\tau$ (on the subspace $\tau_1 = \tau = \tau_2$) it is of interest to explore switching behaviour for $\tau_{12}^0 \neq \tau_{21}^0$ with both values high.  An example of behaviour that can emerge in this parameter regime is shown in Fig.~\ref{Fig:N=2Relaxation}.  Here, we report the triple $(\text{ISI},\tau_1,\tau_2)$ and colour code points according to the value of the instantaneous phase.  We see a loose loop of data points that switches between low and high ISIs in a repetitive fashion.  
This is the numerical signature of the relaxation-type scenario described in Sec.~\ref{Sec:perspective}: slow drift along one attracting branch is followed by a faster transition to another branch, after which the slow drift proceeds in a different direction.

The mechanism for generating such switching behaviour seems to be most robust when the thresholds $h_{ij}$ are centred around values in the vicinity of a commensurate state with $a_{ij} = 2 \pi p_{ij}$ with $p_{ij}\in\mathbb Z_{\ge 0}$.  With such a choice there is a good opportunity for $a_{ij}(t) - h_{ij}$ to change sign and reverse the direction of the slow flow after a jump between fast solution branches, providing a natural route to switching and long-time alternation between distinct phase-locked states rather than settling directly to a single compatible fixed point.  The simulation in Fig.~\ref{Fig:N=2Relaxation} should therefore be interpreted not as a generic outcome for all parameters, but as an illustration of how the nonlinear plasticity rule can organise recurrent switching when the thresholds place the system near competing commensurability classes.

Together, these examples show how the static-delay bifurcation diagrams continue to organise the adaptive system.  When the slow drift is directed toward a compatible fixed point, the dynamics converges to a stationary phase-locked state.  When nearby branches induce opposing slow drifts and the fast subsystem allows transitions between them, the same plasticity rule can generate long-time alternation between distinct phase-locked states.  This motivates the next subsection, where the thresholds in the plasticity rule are used more deliberately to bias a larger network toward commensurate conduction delays.


\subsection{Sculpting a commensurate state}

For an arbitrary network with heterogeneous choices of fixed weights $w_{ij}$ and delays $\tau_{ij}$, one would not expect the spontaneous emergence of a frequency locked state in which all nodes share a common firing frequency. Nevertheless, the white matter plasticity rule introduced above provides a natural mechanism for biasing the system toward a commensurate delay structure. The key observation is that the activity variable $a_{ij}$ associated with the directed edge from node $j$ to node $i$ samples the firing activity of the presynaptic node $j$ over a time window of length $\tau_{ij}$. Thus the relevant clock for choosing a commensurability class on this edge should be the intrinsic or emergent period of the source node $j$, rather than a single externally imposed network period.

This suggests a more general sculpting principle. Rather than trying to impose a global rhythm directly, the adaptive rule can be used locally to make each axonal pathway sensitive to whether its conduction delay spans an integer number of cycles of the activity arriving from its source. In biological terms, the threshold on a given fibre should reflect the typical amount of presynaptic activity expected over one, two, or more cycles of the source neuron. In algorithmic terms, the network is first allowed to reveal the characteristic timescale of each node, and the plasticity thresholds are then chosen so that each edge is most responsive near a delay that is commensurate with the timescale of the node that drives it. The aim is not to prescribe a final phase relationship, but to bias the admissible delay landscape so that mutually compatible phase-locked states become easier for the fast dynamics to select.

One concrete realisation of this principle is to estimate a node-wise target period from the typical inter-spike interval of each source node. For example, we may run the frozen-delay network, or use an early weakly adapted epoch, and record the firing times of each node. We define
\begin{equation}
\Delta_j = \operatorname{median}_m \Delta_j^m, \qquad \Delta_j^m = T_j^{m+1} - T_j^m .
\end{equation}
For each directed edge $(i,j)$ we then assign an integer commensurability class by comparing the initial delay on that edge with the target period of the presynaptic node,
\begin{equation}
p_{ij} = \operatorname{round} \left( \frac{\tau_{ij}(0)}{\Delta_j} \right) = \operatorname{round} \left( \frac{d_{ij}}{c_{ij}(0)\Delta_j} \right).
\end{equation}
We also choose the baseline conduction speed in the plasticity rule to coincide with the initial conduction speed, $c_{ij}^0 = c_{ij}(0)$.
With this convention the unadapted network provides the reference state about which myelination acts. In the absence of activity-dependent modulation the conduction speeds relax back to their initial values, so any subsequent drift in delay space is generated by the plastic component of the rule rather than by an imposed shift of the baseline architecture.
The corresponding threshold is chosen as $h_{ij}=2\pi p_{ij}$.

This choice is motivated by the exact commensurate identity discussed above: if the presynaptic node $j$ is phase-locked with period $\Delta_j$ and the delay satisfies $\tau_{ij}=p_{ij} \Delta_j$, then the activity integral over the propagation window accumulates precisely $p_{ij}$ cycles of firing activity, giving $a_{ij}=2\pi p_{ij}$. Hence $h_{ij}=2\pi p_{ij}$ places the edge close to the switching surface of the $\tanh$ nonlinearity when the delay is near the desired commensurability class.

In this construction the plasticity rule does not prescribe the phase differences between nodes. Rather, it makes each outgoing edge from node $j$ most sensitive when its delay is close to an integer multiple of the typical period of node $j$. The gains $\kappa_{ij}$ and sharpnesses $\sigma_{ij}$ then control how strongly and how abruptly the edge responds to departures from this commensurate configuration. Large $\sigma_{ij}$ makes the plasticity act almost like an edgewise switch, whereas more moderate values provide a smoother and more robust drift in delay space. In this way the adaptive rule sculpts the network toward a set of node-wise commensurate delays, leaving the final phase-locked pattern to be selected by the fast Haken dynamics.

\begin{figure*}[t]
\centering
\begin{minipage}[c]{0.25\textwidth}
\centering
\includegraphics[width=\linewidth]{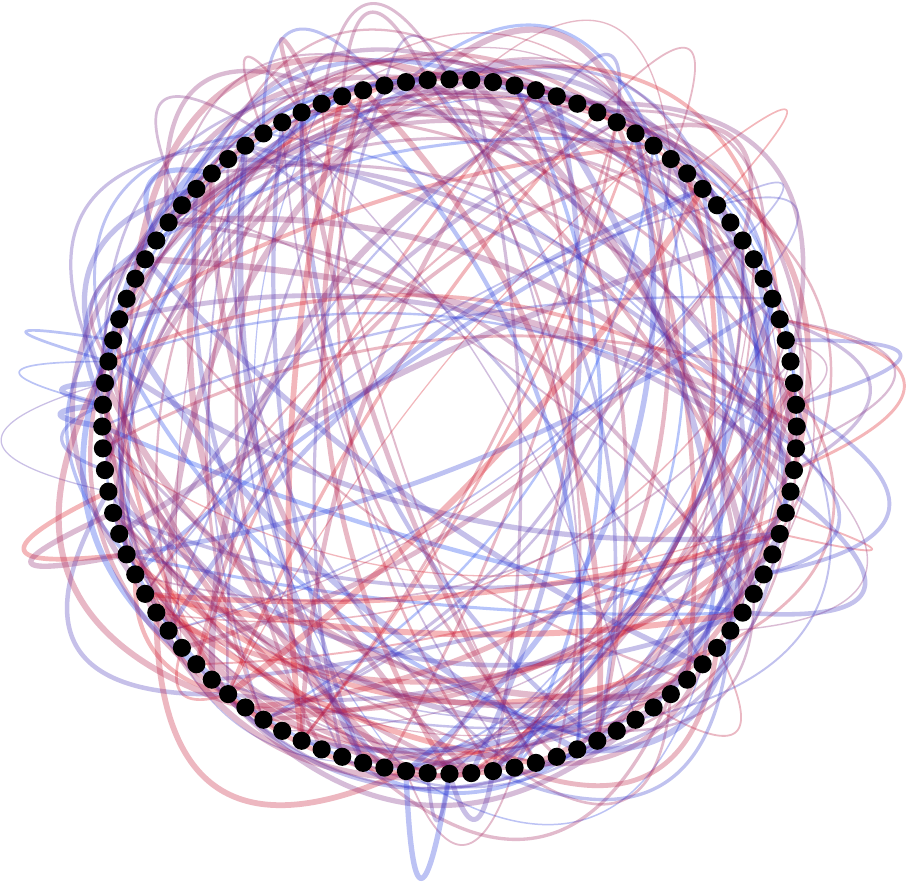}
\end{minipage}
\hfill
\begin{minipage}[c]{0.45\textwidth}
\centering
\includegraphics[width=\linewidth]{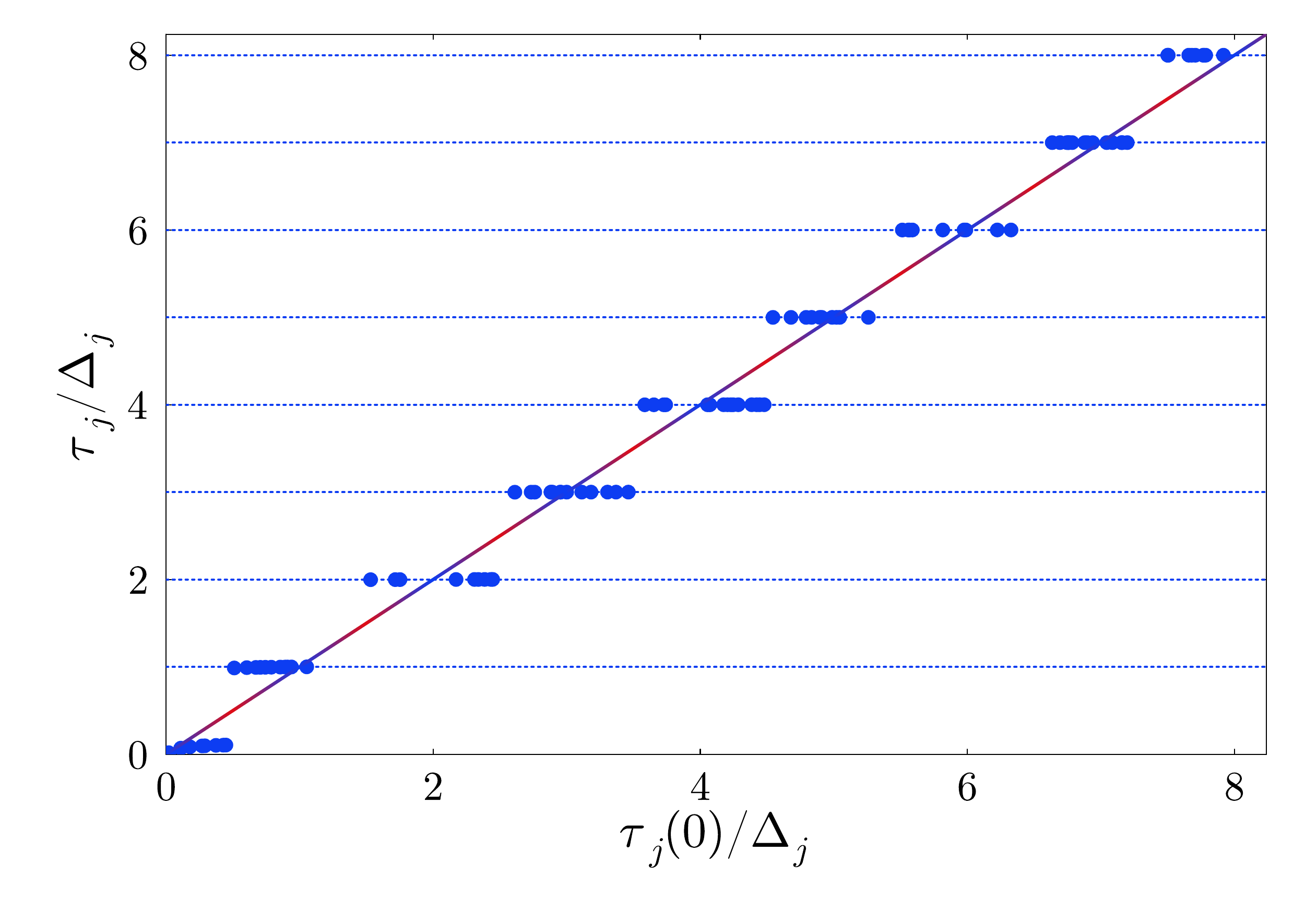}
\end{minipage}
\hfill
\begin{minipage}[c]{0.25\textwidth}
\centering
\includegraphics[width=\linewidth]{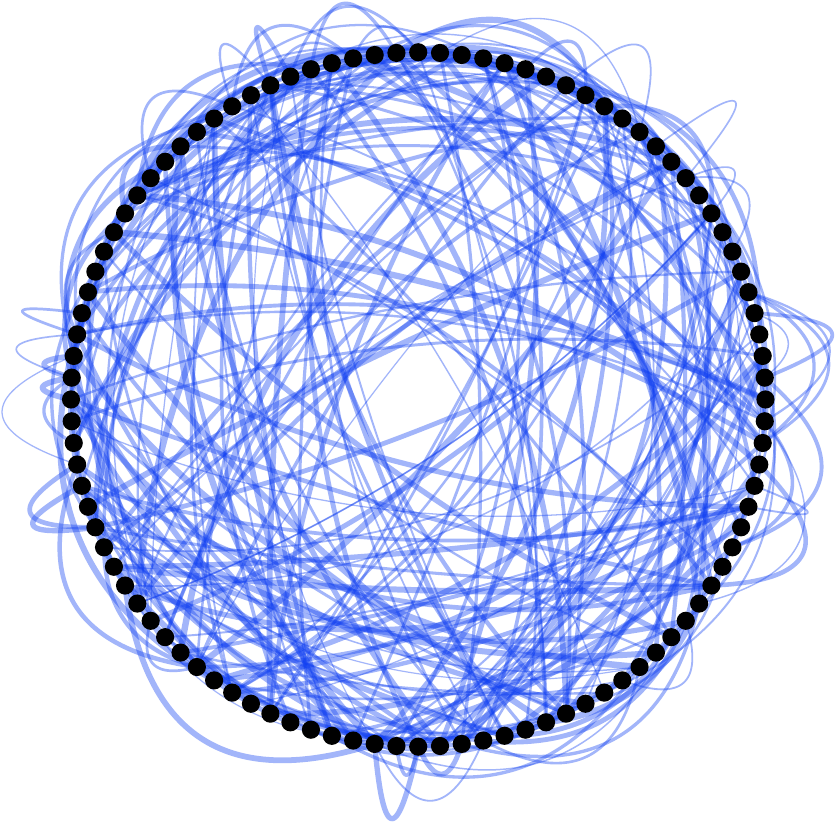}
\end{minipage}
\caption{Schematic and quantification of node-wise commensurability classes for a randomly connected Haken network with $N=100$.  
Left: before adaptation, fibres are coloured by their delay-period mismatch, measured as the distance of $\tau_j(0)/\Delta_j$ from the nearest integer. Blue indicates commensurate delays and red indicates half-integer mismatch. 
Middle: for each source node $j$, a target period $\Delta_j$ is estimated from the median inter-spike interval, and the integer class $p_j=\operatorname{round}(\tau_j(0)/\Delta_j)$ is assigned.  Blue points show the resulting discrete classes, blue horizontal guide lines mark integer values of $p_j$, and the diagonal is coloured by the periodic mismatch map.
Right: after delay sculpting, fibres are coloured by their residual mismatch after assignment to the commensurate class, so that edges lie on blue delay-period classes.  Weights are drawn independently from a uniform distribution on $[-1/N,1/N]$, with self-connections excluded, and initial delays are drawn from a uniform distribution on $(0,100]$.  Parameters are $\alpha=1$, $\beta=10$, $h=0$, $d=1$, $\sigma_j=1$ and $\kappa_j=1$ for all $j$.
\label{Fig:Commensurate}}
\end{figure*}

To reduce the number of state variables we shall consider networks with a fixed distance $d$ between all nodes and a common speed on all edges emanating from node $i$.  In this way we need only consider $N$ distinct speeds with associated delays $\tau_i= d/c_i$.  The white matter plasticity rule (\ref{WM1})-(\ref{WM3}) then reduces to the $N$ (as opposed to $N^2$) equations
\begin{equation}
\begin{split}
\tau_w \FD{}{t} c_i & = c_i^0-c_i + \kappa_i \tanh (\sigma_i (a_i - h_i)), \\
a_i(t) & = \int_{t-\tau_i(t)}^t  \d t' S(\psi_i(t')) , \qquad i=1,\ldots,N .
\end{split}
\end{equation}
Figure~\ref{Fig:Commensurate} shows the result of applying this construction to a randomly connected Haken network.  Here, the weights $w_{ij}$ are quenched random variables drawn from a uniform distribution on $[-1/N,1/N]$.
The horizontal axis shows the initial delay of each source node, normalised by its measured target period, while the vertical axis shows the corresponding integer $p_j$ selected by rounding.  The blue points lie on horizontal integer levels, and the coloured diagonal indicates the delay-period mismatch, measured as the distance of $\tau_j(0)/\Delta_j$ from the nearest integer.  Blue indicates commensurate delays and red indicates half-integer mismatch.  The figure confirms that the rule partitions the heterogeneous set of initial delays into discrete commensurability classes.  These classes do not impose a specific phase-locked state.  Instead, they define the delay values at which the plasticity rule is most sensitive, thereby shaping the slow drift of the network toward conduction delays compatible with the characteristic activity timescales of the source nodes.

From a neuroscientific perspective, a plasticity rule that drives conduction delays into commensurability classes suggests that finite transmission delays need not destroy the phase of an oscillatory signal. If the delay on an outgoing line from node $j$ satisfies $\tau_j \approx p\,T_j$,
for some integer $p$ and characteristic period $T_j$ of the oscillation generated by that node, then the delayed signal returns to approximately the same phase point on the sender's cycle. In this sense, the delay becomes \emph{phase-preserving}: the receiver is driven by an input whose phase still reflects the state of the sending oscillator modulo $2\pi$. Adaptive myelination may therefore be understood not simply as a mechanism for increasing conduction speed, but as a way of tuning transmission delays so that rhythmic phase information is preserved along long-range pathways \cite{Fries2015,Fields2015,Pajevic2014}.
This interpretation is closely aligned with the communication-through-coherence hypothesis, according to which the efficacy of interaction between neuronal populations depends on the phase at which inputs arrive \cite{Fries2015,Schoffelen2005}. A delay that is commensurate with the sender's oscillation period preserves that phase relation up to an integer number of cycles, thereby organising pathways into discrete timing classes. The significance of such a mechanism is not necessarily that it produces a dramatic increase in global synchrony, but that it allows long-range coupling to remain temporally meaningful despite substantial delays. This is consistent with the broader literature on activity-dependent myelination, which suggests that adaptive changes in myelin can regulate conduction velocity, synchrony, and the temporal coordination of distributed circuits \cite{Fields2015,Pajevic2014,Mount2017}, and with experimental evidence that regional myelination can be adjusted so as to preserve timing relationships across pathways of different lengths \cite{Salami2003}. In this view, the main function of a plasticity rule that drives delays into commensurability classes is to render transmission lines approximately phase-transparent, so that oscillatory information can be conveyed without being degraded by the delay itself.

\section{Discussion\label{Sec:Discussion}}

The aim of this paper has been to develop a tractable framework for studying phase-locked states in spiking Haken Lighthouse networks with conduction delays, and then to use this fixed-delay theory as the organising backbone for a model of activity-dependent white matter plasticity.  The starting point is the observation that the Lighthouse model occupies a useful middle ground between detailed spiking descriptions and rate-based neural models: spikes remain the elementary events of communication, but the event-time structure is sufficiently explicit to permit analytical progress.  Building on Haken's original formulation of pulse-coupled neural activity \cite{Haken2002} and the recent revisiting of the Lighthouse model \cite{Coombes2025}, we have shown how multiple delays can be incorporated into the existence and stability theory for phase-locked states.

The first contribution is the fixed-delay phase-locking theory.  For general networks we derived implicit equations for phase-locked states and an associated spectral condition for their linear stability, formulated directly in terms of perturbations of firing times.  This event-time viewpoint is well matched to a spiking model: stability is determined by how small perturbations to one generation of spikes affect subsequent generations.  The approach was illustrated first for a single delayed autapse, where regular-spiking branches, folds and dynamic instabilities can be identified.  In this minimal setting the loss of stability naturally produces modulated inter-spike intervals, providing a mathematically transparent route to jittering of spike trains.  For reciprocal two-node networks, the same formalism separates synchronous, anti-synchronous and asymmetric phase-locked states, and shows how asymmetric branches attach to extrema of symmetric branches.  For ring networks with distance-dependent delays, the circulant structure permits a modal decomposition and gives a direct route to the stability of synchronous, splay and twisted states.

The examples developed here show that conduction delays can act as bifurcation parameters, stability selectors and adaptive state variables, rather than simply as fixed transmission lags.  In neural systems, conduction delays arise from finite axonal propagation speeds and synaptic filtering, and can be comparable to the timescales of rhythmic activity.  In artificial or bio-inspired systems, delayed communication is likewise unavoidable whenever signals propagate through a spatially extended substrate.  The present analysis shows that such delays can shape not only the period of a network rhythm, but also the stability, multiplicity and switching of coherent spiking states.  Thus delay should be regarded as an active component of the network dynamics, rather than  ignored or absorbed into an instantaneous coupling approximation.

The second contribution is to turn this viewpoint into an adaptive theory.  Motivated by activity-dependent myelination, we introduced a phenomenological plasticity rule in which conduction speeds evolve slowly in response to activity over the propagation window of a fibre.  Since the delays are slaved to the conduction speeds, this produces a spiking network with state-dependent delays.  The fixed-delay bifurcation structure then becomes a frozen fast subsystem for a slow--fast adaptive problem.  In this interpretation, the branch diagrams obtained for static delays form an organising skeleton for the full adaptive dynamics: trajectories can drift along stable phase-locked branches, approach slow fixed points, switch between competing branches, or undergo relaxation-type alternations when the slow flow reverses direction between them.


A particularly simple mechanism emerges near commensurate delay--period relationships.  When the propagation time along an edge contains an integer number of cycles of the sending node, the activity accumulated over that propagation window closes over complete periods and no longer depends on the phase at which the window is opened.  The plasticity rule therefore has a natural local reference point: thresholds can be chosen so that each edge is most sensitive near a prescribed integer delay--period class.  In this way, the same local rule can bias different edges toward different commensurate timing relationships.
In the reduced node-wise example of Sec.~7.3, this mechanism partitions outgoing connections into discrete commensurability classes.  Biologically, such a mechanism suggests that adaptive myelination may help preserve the phase of rhythmic signals over long-range connections.  This is naturally related to communication-through-coherence ideas \cite{Fries2015} and to the broader literature on myelin plasticity and neural synchronisation \cite{Pajevic2014,Pajevic2023}.  It also complements recent work on the effect of white matter delays on large-scale brain synchrony, where the delay structure of white matter pathways is treated as a key determinant of collective rhythmic behaviour \cite{Sayli2024}.

From the perspective of network computation, the results suggest a route by which white matter could reshape the attractor landscape of a spiking network.  Phase-locked states provide candidate network states, while delay-induced changes in stability and accessibility provide possible transitions between them.  The adaptive rule studied here does not prescribe a global target state, nor does it guarantee convergence to synchrony.  Instead, it acts locally on edges or outgoing fibres and thereby changes the timing relations that determine which phase-locked states are stable, metastable or reachable.  In this sense, the model supports a cybernetic interpretation of white matter: conduction delays are regulated variables that can shape the timing, stability and transitions of distributed neural activity.  This is consistent with the broader idea that network attractors may be sculpted by modifying the communication delays between oscillatory nodes, rather than only by changing synaptic weights \cite{Kori2001,Ashwin2024}.

There are several limitations to the present study.  The white matter plasticity rule is deliberately phenomenological.  It captures the idea that myelination is activity dependent and slow, but it does not attempt to resolve the cellular mechanisms of axon-glia signalling, sheath formation, internodal length adaptation or metabolic cost \cite{Mount2017,Xin2020,Nishiyama2021,Saab2013,Tepavcevic2021}.  Similarly, each connection is represented by a single effective conduction speed, whereas real white matter pathways contain heterogeneous axons and distributed conduction properties.  The slow--fast interpretation should also be understood as a geometric organising principle rather than a rigorous Fenichel-type invariant-manifold theorem \cite{Fenichel1971}.  The fast subsystem is event driven, and the adaptive model contains state-dependent delays, so the usual hypotheses of smooth finite-dimensional geometric singular perturbation theory do not apply directly.  Finally, most of the examples studied here are low-dimensional or highly structured.  This is intentional: they are chosen to expose mechanisms.  Extending the framework to heterogeneous connectome-like networks will require further analytical and numerical development.

A first direction for future work is therefore to place the slow--fast picture on firmer mathematical foundations.  It would be valuable to develop an invariant-manifold and averaging theory for event-driven spiking systems with slowly evolving state-dependent delays, using the firing-time spectral problem as the replacement for the usual Floquet theory of smooth flows.  A second direction is computational.  Networks with many plastic delays will require continuation tools capable of tracking phase-locked branches, locating folds and dynamic instabilities, and following changes in stability as multiple conduction speeds evolve.  The event-time formulation used here suggests one promising route for such algorithms, especially when combined with modern numerical continuation methods for delay equations \cite{Engelborghs2000,Sieber2014}.  A complementary approach is provided by harmonic balance methods for rate-based neural networks with multiple delays, which have recently been used to construct delay-induced and delay-modulated periodic solutions and their stability in neural mass networks \cite{Coombes2026}.

There are also natural links with phase-amplitude and phase-isostable reductions of oscillator networks.  Such reductions provide a complementary route to understanding how delayed interactions shape dynamics near and away from a limit cycle, including off-cycle perturbations that are important for transitions between network states.  Recent work on phase and amplitude responses for delay equations using harmonic balance \cite{Nicks2024}, and on oscillator network dynamics using a phase-isostable framework \cite{Nicks2024a}, provides a natural basis for extending the present ideas beyond the Lighthouse setting.  One could then compare event-time descriptions of spiking networks with phase-amplitude reductions of more biophysically detailed oscillators and neural mass models.

A further goal is to use adaptive conduction delays to sculpt network attractors.  In symmetric networks, delay tuning may be used to organise heteroclinic switching between unstable phase-locked states.  In heterogeneous or connectome-like networks, where exact symmetry is absent, a more realistic target may be an excitable network attractor in which states are stable but input-dependent transitions occur between them.  The framework developed here provides a starting point for both possibilities: it identifies phase-locked states, determines their stability, and shows how slow delay adaptation can move a system through the corresponding bifurcation structure.  This opens the door to studying sequential memory, robust switching, and input-driven computation in networks whose timing architecture is plastic \cite{Ashwin2024,Rabinovich2018,Maass2002}.

Finally, the same ideas may be useful for studying robustness and degradation.  Since demyelination, ageing and disease can alter conduction speeds, a delay-based theory of phase locking offers a natural language for describing how white matter damage may destabilise coherent activity or alter transitions between network states.  This may be relevant for neurodegenerative and demyelinating diseases, such as multiple sclerosis, where altered conduction timing is expected to disrupt coordinated neural communication \cite{Sorrentino2022}.  Conversely, adaptive myelination may provide a compensatory mechanism for restoring useful timing relationships after perturbation.  Taken together, the results support the view that white matter should be treated not merely as wiring, but as an adaptive dynamical substrate capable of shaping the timing, stability and transitions of spiking network states.

\section*{Acknowledgements}
This work was supported by the Leverhulme Trust through grant RPG-2025-052, ``White Matter Computation: Utilising axonal delays to sculpt network attractors''.

\appendix

\section*{Appendix A\label{Appendix:Branches}}

In this appendix we consider intersections of commensurate asymmetric branches with symmetric branches in the reciprocal $2$-node network.  Here, we show that if an asymmetric branch attaches smoothly to a symmetric branch in the reciprocal $2$-node network of Sec.~\ref{Sec:N=2}, then the point of attachment is necessarily a stationary point of the symmetric branch in the $(\tau,T)$ plane.  

For the reciprocal $2$-node network, the phase-locked states are determined by the pair of equations
\begin{equation}
\mathcal{H}(T,\pm \phi;\tau)=0,
\label{2nodepair_appendix}
\end{equation}
where, from (\ref{H}),
\begin{equation}
\mathcal{H}(T,\phi;\tau)=\int_0^T \d t \, S\left(P(t)+wP(t+\phi T-\tau)\right)-2\pi.
\label{2nodeH}
\end{equation}
Here we have written the dependence on $\tau$ explicitly.  A synchronous branch is defined by
\begin{equation}
\mathcal{H}(T,0;\tau)=0,
\label{syncbranch_appendix}
\end{equation}
while an anti-synchronous branch is defined by $\mathcal{H}\left(T,\frac{1}{2};\tau\right)=0$.

We first consider the case of synchrony.  Suppose that a smooth asymmetric branch intersects the synchronous branch at a point $(T_\ast,\tau_\ast,\phi_\ast)$ with $\phi_\ast=0$.  Along the asymmetric branch we have
\begin{equation}
\mathcal{H}(T,\phi;\tau)=0, \qquad \mathcal{H}(T,-\phi;\tau)=0.
\label{asympair_sync_appendix}
\end{equation}
Subtracting these two equations gives $\mathcal{H}(T,\phi;\tau)-\mathcal{H}(T,-\phi;\tau)=0$.
Using a Taylor expansion around $\phi=0$ as $\mathcal{H}(T,\phi;\tau) = \mathcal{H}(T,0;\tau) + \phi \left. \partial {\mathcal{H}} /\partial \phi \right |_{(T,0;\tau)} + O(\phi^2)$ yields $\mathcal{H}(T,\phi;\tau)-\mathcal{H}(T,-\phi;\tau) = 2 \phi \left. \partial {\mathcal{H}} /\partial \phi \right |_{(T,0;\tau)} + O(\phi^3)$.
Dividing by $2\phi$ and taking the limit $\phi\to 0$ yields
\begin{equation}
\PD{\mathcal{H}}{\phi}(T_\ast,0;\tau_\ast)=0.
\label{Hphi_zero_sync_appendix}
\end{equation}

To relate this to the slope of the synchronous branch, we differentiate (\ref{syncbranch_appendix}) with respect to $\tau$ along a smooth solution $T=T(\tau)$.  This gives
\begin{equation}
\PD{\mathcal{H}}{T}(T,0;\tau) \FD{T}{\tau}+\PD{\mathcal{H}}{\tau}(T,0;\tau)=0.
\label{branchdiff_sync_appendix}
\end{equation}
Hence, provided that
\begin{equation}
\PD{\mathcal{H}}{T}(T_\ast,0;\tau_\ast)\neq 0,
\label{HT_nonzero_sync_appendix}
\end{equation}
we obtain
\begin{equation}
\FD{T}{\tau}(\tau_\ast)=-\frac{\PD{\mathcal{H}}{\tau}(T_\ast,0;\tau_\ast)}{\PD{\mathcal{H}}{T}(T_\ast,0;\tau_\ast)}.
\label{branchslope_sync_appendix}
\end{equation}
Thus it remains to show that $\partial \mathcal{H} /\partial {\tau}=0$ at the point of intersection.
This follows directly from the form of $\mathcal{H}$ in (\ref{2nodeH}).  Introducing $A=\phi T-\tau$,
we see that $\phi$ and $\tau$ enter $\mathcal{H}$ only through the combination $A$.  Differentiating (\ref{2nodeH}) directly with respect to $\tau$ and $\phi$ shows that
\begin{equation}
\PD{\mathcal{H}}{\tau}(T,\phi;\tau)=-\frac{1}{T}\PD{\mathcal{H}}{\phi}(T,\phi;\tau).
\label{HtauHphi_identity_appendix}
\end{equation}
Evaluating (\ref{HtauHphi_identity_appendix}) at the point of attachment and using (\ref{Hphi_zero_sync_appendix}) we obtain
\begin{equation}
\PD{\mathcal{H}}{\tau}(T_\ast,0;\tau_\ast)=0.
\label{Htau_zero_sync_appendix}
\end{equation}
Substitution into (\ref{branchdiff_sync_appendix}) now gives
\begin{equation}
\FD{T}{\tau}(\tau_\ast)=0.
\label{stationary_sync_appendix}
\end{equation}
Hence the synchronous branch is stationary at the point where the asymmetric branch attaches.
The proof for the anti-synchronous branch is identical and we do not present it here.
We have therefore established the following: \textit{if an asymmetric branch attaches smoothly to either the synchronous branch $\phi=0$ or the anti-synchronous branch $\phi=1/2$, then the point of attachment is a stationary point of the corresponding symmetric branch in the $(\tau,T)$ plane, provided that $\partial {\mathcal{H}} / \partial {T}\neq 0$ at that point.}

\bibliographystyle{unsrtnat}
\bibliography{PhaseLockedStates}

@article{Ruschel2026,
	author = {S Ruschel and E Hristov and H G E Meijer and S Coombes and R Nicks},
	doi = {10.48550/arXiv.2605.23520},
	journal = {Submitted},
	title = {Network attractors driven by time-delay plasticity},
	year = {2026}}

@article{Fries2005,
	author = {P Fries},
	journal = {Trends in Cognitive Sciences},
	pages = {474-480},
	title = {A mechanism for cognitive dynamics: neuronal communication through neuronal coherence},
	volume = {9},
	year = {2005}}

@article{Coombes2026,
	author = {S Coombes and H G E Meijer},
	doi = {10.48550/arXiv.2509.22495},
	journal = {Mathematical Neuroscience and its Applications},
	note = {Under revision},
	title = {Synchrony in firing rate neural networks with multiple delays: A harmonic balance approach},
	year = {2026}}

@article{Sorrentino2022,
	author = {Sorrentino, P. and Rucco, R. and Lardone, A. and Liparoti, M. and Trojsi, F. and Curcio, G. and Baselice, F. and Sorrentino, G. and Mandolesi, L. and Babiloni, C.},
	journal = {Journal of Neuroscience},
	pages = {8807-8816},
	title = {Whole-brain propagation delays in multiple sclerosis, a combined tractography-magnetoencephalography study},
	volume = {42},
	year = {2022}}

@article{Maass2002,
	author = {W Maass and T Natschl{\"a}ger and H Markram},
	journal = {Neural Computation},
	number = {11},
	pages = {2531-2560},
	title = {Real-time computing without stable states: {A} new framework for neural computation based on perturbations},
	volume = {14},
	year = {2002}}

@article{Rabinovich2018,
	author = {M I Rabinovich and P Varona and A I Selverston and H D I Abarbanel},
	journal = {Reviews of Modern Physics},
	pages = {1213-1265},
	title = {Dynamical principles in neuroscience},
	volume = {78},
	year = {2006}}

@article{Engelborghs2000,
	author = {K Engelborghs and T Luzyanina and D Roose},
	journal = {Journal of Computational and Applied Mathematics},
	pages = {265275},
	title = {Numerical bifurcation analysis of delay differential equations},
	volume = {125},
	year = {2000}}

@article{Sieber2014,
	author = {J Sieber and K Engelborghs and T Luzyanina and G Samaey and D Roose},
	journal = {arXiv preprint arXiv:1406.7144},
	title = {{DDE-BIFTOOL manual: Bifurcation analysis of delay differential equations}},
	year = {2014}}

@article{Fenichel1971,
	author = {N Fenichel},
	journal = {Indiana University Mathematics Journal},
	pages = {193-226},
	title = {Persistence and smoothness of invariant manifolds for flows},
	volume = {21},
	year = {1971}}

@article{Nishiyama2021,
	author = {A Nishiyama and R Suzuki and X Zhu},
	journal = {Seminars in Cell \& Developmental Biology},
	pages = {25-37},
	title = {Life-long oligodendrocyte development and plasticity},
	volume = {116},
	year = {2021}}

@article{Saab2013,
	author = {A S Saab and I D Tzvetanova and K A Nave},
	journal = {Current Opinion in Neurobiology},
	pages = {1065-1072},
	title = {The role of myelin and oligodendrocytes in axonal energy metabolism},
	volume = {23},
	year = {2013}}

@article{Tepavcevic2021,
	author = {V Tepav{\v{c}}evi\'{c}},
	journal = {Life},
	number = {3},
	pages = {238},
	title = {Oligodendroglial energy metabolism and (re)myelination},
	volume = {11},
	year = {2021}}

@article{Ashwin2024,
	author = {P Ashwin and M Fadera and C Postlethwaite},
	journal = {Current Opinion in Neurobiology},
	pages = {102818},
	title = {{Network attractors and nonlinear dynamics of neural computation}},
	volume = {84},
	year = {2024}}

@article{Kori2001,
	author = {Kori, H and Kuramoto, Y},
	journal = {Physical Review E},
	pages = {046214},
	title = {{Slow switching in globally coupled oscillators: robustness and occurrence through delayed coupling}},
	volume = {63},
	year = {2001}}

@article{Sayli2024,
	author = {M {\c{S}ayli} and S Coombes},
	journal = {Communications in Nonlinear Science and Numerical Simulation},
	pages = {107803},
	title = {Understanding the effect of white matter delays on large scale brain synchrony},
	volume = {131},
	year = {2024}}

@article{Schoffelen2005,
	author = {J-M Schoffelen and R Oostenveld and P Fries},
	journal = {Science},
	pages = {111-113},
	title = {Neuronal coherence as a mechanism of effective corticospinal interaction},
	volume = {308},
	year = {2005}}

@article{Fields2015,
	author = {R D Fields},
	journal = {Nature Reviews Neuroscience},
	pages = {756-767},
	title = {A New Mechanism of Nervous System Plasticity: Activity-Dependent Myelination},
	volume = {16},
	year = {2015}}

@article{Salami2003,
	author = {M Salami and C Itami and T Tsumoto and F Kimura},
	journal = {Proceedings of the National Academy of Sciences USA},
	pages = {6174-6179},
	title = {Change of Conduction Velocity by Regional Myelination Yields Constant Latency Irrespective of Distance between Thalamus and Cortex},
	volume = {100},
	year = {2003}}

@article{Pinsky1995,
	author = {P F Pinsky and J Rinzel},
	journal = {Biological Cybernetics},
	pages = {129-137},
	title = {Synchrony measures for biological neural networks},
	volume = {73},
	year = {1995}}

@article{Gansel2022,
	author = {K S Gansel},
	journal = {Frontiers in Integrative Neuroscience},
	pages = {900715},
	title = {Neural synchrony in cortical networks: mechanisms and implications for neural information processing and coding},
	volume = {16},
	year = {2022}}

@article{Xin2020,
	author = {W Xin and J R Chan},
	journal = {Nature Reviews Neuroscience},
	pages = {682-694},
	title = {Myelin plasticity: sculpting circuits in learning and memory},
	volume = {21},
	year = {2020}}

@article{Mount2017,
	author = {C W Mount and M Monje},
	journal = {Neuron},
	pages = {743-756},
	title = {Wrapped to adapt: experience-dependent myelination},
	volume = {95},
	year = {2017}}

@article{Campbell2007,
	author = {S A Campbell},
	journal = {Handbook of brain connectivity},
	pages = {65-90},
	title = {Time delays in neural systems},
	year = {2007}}

@incollection{Jones1995,
	address = {Berlin, Heidelberg},
	author = {C K R T Jones},
	booktitle = {Dynamical Systems},
	editor = {R Johnson},
	pages = {44--118},
	publisher = {Springer},
	series = {Lecture Notes in Mathematics},
	title = {Geometric Singular Perturbation Theory},
	volume = {1609},
	year = {1995}}

@article{Klinshov2024,
	author = {V V Klinshov and V I Nekorkin},
	journal = {Chaos: An Interdisciplinary Journal of Nonlinear Science},
	pages = {033101},
	title = {Adaptive myelination causes slow oscillations in recurrent neural loops},
	volume = {34},
	year = {2024}}

@article{Foss1996,
	author = {J Foss and A Longtin and B Mensour and J Milton},
	journal = {Physical Review Letters},
	pages = {708-711},
	title = {Multistability and Delayed Recurrent Loops},
	volume = {76},
	year = {1996}}

@article{Klinshov2015,
	author = {V Klinshov and L L\"ucken and D Shchapin and V Nekorkin and S Yanchuk},
	journal = {Physical Review Letters},
	pages = {178103},
	title = {Multistable Jittering in Oscillators with Pulsatile Delayed Feedback},
	volume = {114},
	year = {2015}}

@article{Coombes2012,
	author = {S Coombes and R Thul and K C A Wedgwood},
	journal = {Physica D},
	pages = {2042--2057},
	title = {Nonsmooth dynamics in spiking neuron models},
	volume = {241},
	year = {2012}}

@article{Coombes2025,
	author = {S Coombes},
	journal = {The European Physical Journal Special Topics : In Memoriam Hermann Haken: Synergetics and Self-organisation in Complex Systems},
	title = {{Revisiting the Haken Lighthouse Model}},
	url = {https://doi.org/10.1140/epjs/s11734-025-01841-3},
	year = {2025}}

@article{DiCaterina2024,
	author = {G {Di Caterina} and M Zhang and J Liu},
	journal = {Frontiers in Neuroscience},
	pages = {1406502},
	title = {Editorial: Theoretical advances and practical applications of spiking neural networks},
	volume = {18},
	year = {2024}}

@book{Haken2002,
	address = {Berlin, Heidelberg},
	author = {H Haken},
	publisher = {Springer-Verlag},
	title = {Brain Dynamics: Synchronization and Activity Patterns in Pulse-coupled Neural Nets with Delays and Noise},
	year = {2002}}

@article{Jolly2026,
	author = {G Jolly and R Nicks and S Ruschel and G Iskenderoglu and S Coombes},
	journal = {Chaos},
	title = {Phase oscillator networks with multiple and state-dependent delays: A framework for exploring white matter plasticity in neurodynamics},
	volume = {In prep},
	year = {2026}}

@article{Ornes2025,
	author = {S Ornes},
	journal = {Proceedings of the National Academy of Sciences USA},
	pages = {{e2528654122}},
	title = {{Can neuromorphic computing help reduce AI's high energy cost?}},
	volume = {122},
	year = {2025}}

@article{Tavakoli2024,
	author = {S K Tavakoli and A Longtin},
	journal = {Physical Review E},
	pages = {054203},
	title = {Boosting reservoir computer performance with multiple delays},
	volume = {109},
	year = {2024}}

@article{Nicks2024a,
	author = {R Nicks and R Allen and S Coombes},
	journal = {Chaos},
	pages = {013141},
	title = {Insights into oscillator network dynamics using a phase-isostable framework},
	volume = {34},
	year = {2024}}

@article{Huber2018,
	author = {E Huber and P M Donnelly and A Rokem and J D Yeatman},
	journal = {Nature Communications},
	pages = {2260},
	title = {Rapid and widespread white matter plasticity during an intensive reading intervention},
	volume = {9},
	year = {2018}}

@article{Fries2015,
	author = {P Fries},
	journal = {Neuron},
	pages = {220-235},
	title = {Communication Through Coherence {(CTC 2.0)}},
	volume = {88},
	year = {2015}}

@article{Nicks2024,
	author = {R Nicks and R Allen and S Coombes},
	journal = {Physical Review E},
	pages = {L012202},
	title = {Phase and amplitude responses for delay equations using harmonic balance},
	volume = {110},
	year = {2024}}

@article{Pajevic2023,
	author = {S Pajevic and D Plenz and P J Basser and R D Fields},
	journal = {eLife},
	pages = {e81982},
	title = {Oligodendrocyte-mediated myelin plasticity and its role in neural synchronization},
	volume = {12},
	year = {2023}}

@article{Pajevic2014,
	author = {S Pajevic and P J Basser and R D Fields},
	journal = {Neuroscience},
	pages = {135-147},
	title = {{Role of myelin plasticity in oscillations and synchrony of neuronal activity}},
	volume = {276},
	year = {2014}}

@article{Talidou2022,
	author = {A Talidou and P W Frankland and D Mabbott and J Lefebvre},
	journal = {Nature Computational Science},
	pages = {665-676},
	title = {{Homeostatic coordination and up-regulation of neural activity by activity-dependent myelination}},
	volume = {2},
	year = {2022}}

@article{Talidou2021,
	author = {A Talidou and P W Frankland and D Mabbott and J Lefebvre},
	journal = {bioRxiv},
	title = {Learning to be on time: temporal coordination of neural dynamics by activity-dependent myelination},
	year = {2021}}

@article{Noori2020,
	author = {R Noori and D Park and J D Griffiths and S Bells and P W Frankland and D Mabbott and J Lefebvre},
	journal = {Proceedings of the National Academy of Sciences USA},
	pages = {13227-13237},
	title = {{Activity-dependent myelination: A glial mechanism of oscillatory self-organization in large-scale brain networks}},
	volume = {117},
	year = {2020}}

@article{Vivo2019,
	author = {L de Vivo and M Bellesi},
	journal = {Glia},
	pages = {2142-2152},
	publisher = {Wiley Online Library},
	title = {The role of sleep and wakefulness in myelin plasticity},
	volume = {67},
	year = {2019}}

@article{Gibson2014,
	author = {E M Gibson and D Purger and C W Mount and A K Goldstein and G L Lin and L S Wood and I Inema and S E Miller and G Bieri and J B Zuchero and B A Barres and P J Woo and H Vogel and M Monje},
	journal = {Science},
	pages = {1252304},
	title = {Neuronal Activity Promotes Oligodendrogenesis and Adaptive Myelination in the Mammalian Brain},
	volume = {344},
	year = {2014}}

@article{Fields2008,
	author = {Fields, R Douglas},
	journal = {Trends in Neurosciences},
	pages = {361-370},
	publisher = {Elsevier},
	title = {White matter in learning, cognition and psychiatric disorders},
	volume = {31},
	year = {2008}}

@article{Lefebvre2025,
	author = {J Lefebvre and A Clappison and A Longtin and A Hutt},
	journal = {Communications Physics},
	number = {1},
	pages = {145},
	title = {{Myelin-induced gain control in nonlinear neural networks}},
	volume = {8},
	year = {2025}}

@article{Karimian2019,
	author = {M Karimian and D Dibenedetto and M Moerel and T Burwick and R L Westra and P De Weerd and M Senden},
	journal = {Chaos},
	pages = {083122},
	title = {Effects of synaptic and myelin plasticity on learning in a network of {K}uramoto phase oscillators},
	volume = {29},
	year = {2019}}

@article{Park2020,
	author = {S H Park and J Lefebvre},
	journal = {The Journal of Mathematical Neuroscience},
	number = {1},
	pages = {16},
	title = {{Synchronization and resilience in the Kuramoto white matter network model with adaptive state-dependent delays}},
	volume = {10},
	year = {2020}}

\end{document}